\pdfoutput=1
\documentclass[11pt]{article}

\usepackage[]{acl}
\usepackage{graphicx}
\usepackage{enumitem}
\usepackage{hyperref}
\usepackage{times}
\usepackage{latexsym}
\usepackage{bm}
\usepackage[T1]{fontenc}
\usepackage[utf8]{inputenc}
\usepackage{booktabs}
\usepackage{multirow}
\usepackage{subfig}
\usepackage{colortbl}
\usepackage[table]{xcolor}
\usepackage{lipsum}
\usepackage{longtable}
\usepackage{makecell}

\usepackage{listings}
\usepackage{placeins}
\usepackage{siunitx}

\usepackage{url}
\usepackage{xurl}

\newcommand{\appref}[1]{\hyperref[#1]{Appendix~\ref*{#1}}}

\usepackage{tcolorbox}
\newtcolorbox{promptbox}[1][]{
  colback=gray!8,
  colframe=gray!50,
  fonttitle=\bfseries\small,
  title=#1,
  left=6pt, right=6pt, top=4pt, bottom=4pt,
  fontupper=\small\ttfamily
}

\tcbuselibrary{breakable}
\newtcolorbox{promptboxLONG}[1][]{
  breakable,
  colback=gray!8,
  colframe=gray!50,
  fonttitle=\bfseries\small,
  title=#1,
  left=6pt, right=6pt, top=4pt, bottom=4pt,
  fontupper=\small\ttfamily
}

\usepackage{microtype}

\usepackage{arydshln}
\usepackage{amsmath}

\definecolor{lightyellow}{RGB}{255, 255, 204}

\newlength{\bibitemsep}
\newlength{\bibparskip}

\newcommand{\email}[1]{\href{mailto:#1}{#1}}

\title{AI-Assisted Peer Review Across Research Communities: From Reviewer AI Policies to LLM Review Quality}

\author{Alexander M. Fichtl, {\bf Lukas Ellinger}, {\bf Josefin Kelber}, {\bf Kryštof Olík} \and {\bf Georg Groh} \\
         Social Computing Research Group \\Technical University of Munich\\Boltzmannstraße 3, 85748, Garching, Germany\\\email{\{alexander.fichtl, lukas.ellinger, josefin.kelber, krystof.olik, georg.groh\}@tum.de}}
         
\begin{document}
\maketitle

\begin{abstract}
AI-assisted peer review is increasingly discussed and adopted as a tool to support the scientific publishing process, yet there is little systematic understanding of how publication venues regulate its use or of how capable current AI review systems are. We address these questions by first surveying reviewer-facing AI policies across 111 leading AI/NLP conferences and medical journals, revealing substantial regulation differences between the two communities. Second, we evaluate AI-generated peer reviews at ICLR 2026 and Nature Communications using a novel dataset comprising original manuscript submissions and several hundred human- and machine-generated reviews. We compare reviews produced by open-source and proprietary models using complementary evaluation metrics, including LLM-as-a-Judge, score alignment, granularity, and overlap with human reviewers' concerns. Our results show that current LLMs can generate detailed reviews and extend critique coverage, but exhibit weaknesses, such as overly positive recommendations, generic criticism, and uneven evidence grounding. We demonstrate that aggregate quality scores alone can overestimate review quality and argue for multi-dimensional evaluation of AI-generated peer reviews.
\end{abstract}

\section{Introduction}\label{section:introduction}
Peer review is central to scientific publishing, providing quality control, editorial guidance, and feedback to authors. However, the process is affected by rapidly rising submission volumes \citep{kim2025position}, reviewer burden, delays, variability in review quality, and disagreement among reviewers \citep{biswas2026aaai, liang2024can}. Recent progress in large language models (LLMs) has increased interest in AI-assisted peer reviewing systems that may support reviewers, editors, or authors \citep{kuznetsov2024can, acl2026_keynote_zhang}. In practice, these tools are already in use: a 2025 survey reports that 53\% of reviewers use AI during peer review \citep{frontiers2025whitepaper}, and text-analysis estimates that 6.5-16.9\% of review text at major venues show LLM usage \citep{liang2024monitoring}. Every main-track submission at AAAI 2026 received one clearly identified AI review from a state-of-the-art system, making it the first large-scale field deployment of AI-assisted peer review \citep{biswas2026aaai}. Moreover, in the ARR May 2026 cycle, the reviewer support tool REVAS\footnote{\href{http://aclrollingreview.org/revas-may2}{ARR statement}, \href{https://revas.mbzuai.ac.ae/}{REVAS website}} was introduced on a trial basis in response to AI usage. 

Nevertheless, it remains unclear whether LLMs can produce critical, grounded reviews aligned with human expert judgment, and evaluating them is challenging because peer review formats vary across venues and are inherently open-ended: multiple valid reviews may emphasize different aspects of the same manuscript. A fair evaluation also requires access to the manuscript version that reviewers actually saw, since later revisions may already address reviewer concerns. In this paper, we first survey reviewer-facing AI policies across 111 leading AI/NLP conferences and medical journals, then propose a dataset of original submissions to ICLR 2026 and Nature Communications, along with human- and AI-generated reviews. Reviews are generated using three multimodal LLMs: Llama-4, Qwen3, and GPT-5. For ICLR, we use a multi-step prompting strategy that generates each review field separately; for Nature Communications, we use a single-pass prompt that produces a narrative reviewer report. The chosen venues offer complementary settings: ICLR provides public, structured reviews and numerical scores via OpenReview, while Nature Communications offers journal-style, transparent peer-review reports. Finally, we evaluate the quality of the generated reviews through complementary evaluation metrics.

Our results show that LLM-generated reviews can receive high aggregate evaluation scores while exhibiting systematic weaknesses in specificity, grounding in manuscript evidence, and producing overly positive assessments. Their critiques do not consistently align with substantive human concerns; however, they can provide additional coverage. These findings underscore the need to evaluate AI-assisted reviewing systems using complementary metrics and to retain human authority over consequential review decisions.

The main contributions of this paper are:

\begin{itemize}[noitemsep, topsep=2pt]
    \item A survey of AI and medical conferences and journals regarding their use of AI-assisted reviewing, including a categorization of the observed forms of adoption.
    \item A dataset containing original paper submissions and several hundred human and AI-generated reviews from ICLR 2026 and Nature Communications.
    \item An evaluation suite combining LLM-as-a-judge scoring, Granuscore, and an overlap metric based on concern decomposition and paper-grounded LLM matching against an in-corpus human baseline.
    \item We provide our venue-specific prompting setup for generating structured conference reviews and narrative journal-style reviews, and release our code and data on \href{https://github.com/alexander-fichtl/ai_assisted_peer_review}{GitHub}.
\end{itemize}

\section{Related Work}\label{section:related_work}

\paragraph{Surveys on Publisher AI-use}

\citet{zhuang2025large} categorize publisher-level AI policy positions among major publishers, including Elsevier, Springer Nature, IEEE, ACM, Frontiers, and Wiley. The authors state that “most of the publishers prohibit reviewers from using AIGC tools to generate
or assist in writing review reports” for two stated reasons: confidentiality risk and reviewer accountability. They also state that academia “exhibits a more tolerant stance, accepting their use in reviewing under certain conditions.”  \citet{mollaki2024death} argues that author-side AI policies have already been established across publishers, but reviewer-side AI policies are still missing or are inconsistent: only two out of ten publishers mention AI for peer review in their policies, against the broader author-side coverage \citet{zhuang2025large} records. They also state that written policies alone are not enough, and that there is a need for transparent processes to detect noncompliance, investigation, and exclude reviewers found to have used AI in their reviews without disclosure. 

While these two surveys are related to our work, the policy landscape of AI-assisted review has, to the best of our knowledge, not yet been systematically surveyed at the venue level, a gap that we are aiming to close.

\paragraph{Datasets}

\citet{gonzalez2024learning} published a dataset compiled from a complete scrape of ICLR submissions from OpenReview. The current version (26v1) contains 55,906 ICLR submissions from 2017 to 2026. However, the provided submissions are revised versions that have already incorporated the review feedback, making them much less valuable for testing an AI review-generation pipeline. Generating AI reviews for papers that have already been reviewed and updated makes little sense and also hinders evaluation against the in-corpus human review "ground truth".
For Nature Communications, the MMSCI dataset \citep{li2024mmsci} was collected from the Nature Communications website and comprises papers across five major categories and 72 subjects. These have, however, also incorporated reviewer feedback already. For other venues and sources, we refer readers to \citet{kuznetsov2024can}, who proposed a companion repository listing existing peer-review datasets with brief descriptions for each.

\paragraph{Review Generation}
\citet{tyser2024ai} proposed a comprehensive bias-aware pipeline for review
generation that inspired our approach. \citet{xu2025llms} additionally
incorporate retrieval of related papers, enabling models to identify missing
baselines and citations. Beyond individual methods, large venues have started
exploring AI-assisted review pipelines; for example, AAAI 2026 deployed an AI
review pipeline for all main-track submissions \citep{biswas2026aaai}. Prior
work has also identified limitations of AI-generated reviews, including a
tendency to overrate borderline papers \citep{latona2024lottery} and
vulnerability to hidden instructions in submissions \citep{lin2025hidden}.
For a more comprehensive overview of methods for generating reviews, we refer to the survey by \citet{survey_of_ai_review_methods}.

\paragraph{Evaluation}
\citet{du2024llms} compare human and AI reviews sentence by sentence. They found that even SOTA LLMs write significantly more unhelpful or flawed sentences than human reviewers do, aligning with \citet{robertson2023gpt4} who found GPT-4 only ``slightly helpful'' as a reviewer, reliable on obvious issues but generic on deeper critique. \citet{sun2025llm} conclude in their survey that current LLMs are review aids rather than autonomous reviewers. Our own evaluation approach is closest to \citet{liang2024can}, who ran an AI review pipeline with GPT-4 on 3{,}096 Nature-family papers and 1{,}709 ICLR submissions and measured LLM-vs-human feedback overlap of 30.85\% (Nature) and 39.23\% (ICLR), which is very close to human-vs-human overlap on the same corpus. We extend this line of work by evaluating multiple LLMs across conference and journal settings using complementary metrics that capture score alignment, concern overlap, granularity, and overall review quality.

\setlength{\belowcaptionskip}{-10pt}
\section{Survey of AI Review Policies}
\begin{figure*}[t]
  \centering
  \subfloat{%
    \includegraphics[width=0.49\textwidth]{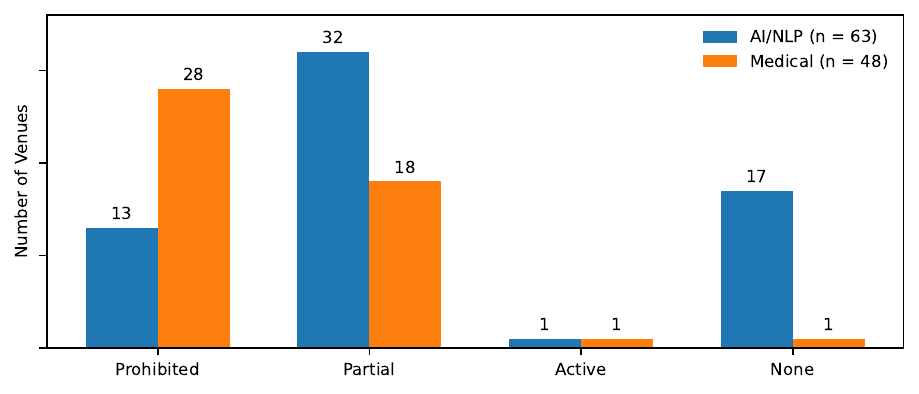}}
  \hfill
  \subfloat{%
    \includegraphics[width=0.49\textwidth]{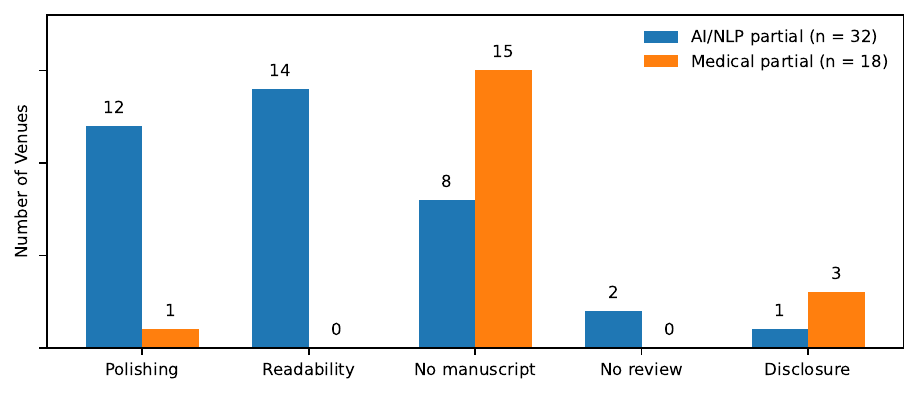}}
    \caption{Reviewer-AI policies across 63 AI/NLP conferences and 48 medical journals. Left: top-level policy categories. Right: non-exclusive subtags among Partial venues.}
  \label{fig:survey_distributions}
\end{figure*}

As the use of AI during peer review becomes increasingly common, publication venues have begun to introduce policies that govern its use. However, these policies vary across research communities and have not yet been systematically compared at the venue level. To provide a cross-disciplinary perspective, we survey current reviewer-facing AI policies across leading AI/NLP conferences and medical journals.

\subsection{Venue Collection}
We survey 111 venues: 63 AI/NLP conferences and 48 medical journals. AI/NLP venues are selected from the \href{https://portal.core.edu.au/conf-ranks/}{CORE Rankings Portal} by including all A* and A-ranked conferences whose Field-of-Research codes fall under Artificial Intelligence, Machine Learning, Computer Vision, or Data Management and Data Science. Medical journals are selected from the \href{https://clarivate.com/academia-government/scientific-and-academic-research/research-funding-analytics/journal-citation-reports}{Clarivate Journal Citation Reports} by selecting the eight highest-impact ones (2024 Journal Impact Factor) from six Clinical Medicine categories: Medicine General \& Internal, Oncology, Cardiac \& Cardiovascular Systems, Clinical Neurology, Radiology, Nuclear Medicine \& Medical Imaging, and Health Care Sciences \& Services.

Between 1 and 27 May 2026, we manually collected reviewer-facing AI policies from official venue or publisher websites (if inherited) and recorded relevant policy excerpts and URLs.

\subsection{Policy Taxonomy}
Each venue is assigned to one of four mutually exclusive categories: \textbf{Prohibited}, \textbf{Partial}, \textbf{Active}, and \textbf{No policy}. Partial policies receive additional non-exclusive tags capturing permitted uses (grammar polishing or improving the readability of the review) and restrictions (no manuscript upload, no review upload, or mandatory disclosure of AI use). This allows us to compare not only the overall policy landscape but also the specific forms of AI use permitted across venues.

\subsection{Survey Findings}
\autoref{fig:survey_distributions} shows substantially different policy landscapes across communities. Among AI/NLP conferences, Partial policies are most common (32/63, 51\%), followed by No policy (17/63, 27\%) and Prohibited (13/63, 21\%). AAAI 2026 is the only venue in this group with an active AI-assisted review pipeline. Medical journals show the opposite trend: most prohibit reviewer AI use (28/48, 58\%), while 18 (38\%) partially allow it and only one has no policy. This difference is consistent with the more developed publication-ethics infrastructure in medicine, particularly the ICMJE Recommendations to which most medical journals subscribe. Similar to AI/NLP, only one venue in our sample actively employs AI-assisted reviewing, namely IEEE Transactions on Medical Imaging through its AI4TMI initiative.

Partial policies also differ: AI/NLP venues mainly permit language assistance, whereas medical journals primarily impose confidentiality restrictions, with 15/18 prohibiting manuscript uploads.

A complete venue-by-venue classification is provided in \appref{app:survey-tables}.

\section{Experimental Setup}
\label{section:methodology}
While the previous section examined how publication venues regulate AI-assisted reviewing, this section evaluates the capabilities of current LLMs to generate peer reviews for scientific papers. We first describe the datasets, followed by the review-generation pipeline and the evaluation metrics.

\subsection{Datasets}
We evaluate AI-generated peer reviews on two complementary publication venues: \href{https://iclr.cc}{ICLR 2026} and \href{https://www.nature.com/ncomms/}{Nature Communications}. We selected ICLR because it is widely used as a data source in prior work on automated peer review \citep{liang2024can, tyser2024ai, latona2024lottery, xu2025llms}, and because its review process is publicly accessible through \href{https://openreview.net}{OpenReview}. We include Nature Communications to evaluate journal-style peer review, which differs in its review process, enabling a comparison across both publication settings.

\subsubsection{ICLR 2026}
ICLR exposes review histories through OpenReview. However, it does not preserve the original PDFs at the time of initial submission viewed by reviewers. Consequently, generating reviews from the currently available paper versions may underestimate reviewer concerns that were already addressed. Existing datasets, such as \citet{gonzalez2024learning}, do not consistently provide the original submissions. We therefore collected the ICLR 2026 submissions after editorial decisions were made public but before revised versions appeared on OpenReview, capturing the manuscripts in the state reviewers saw.

Our dataset comprises 50 randomly sampled submissions: 29 rejected, 20 poster, and one oral acceptance. Papers (including the appendix) contain $15{,}334 \pm 5{,}805$ words and $17.3 \pm 14.7$ figures on average and received $3.8 \pm 0.5$ reviews.

\subsubsection{Nature Communications}
Unlike ICLR, Nature Communications publishes continuously. We consider papers accepted in 2024 and 2025 that have transparent peer-review records. We sourced original manuscripts from \href{https://www.researchsquare.com}{Research Square}, where authors may upload their paper upon submission, preserving the version evaluated by reviewers. We were unable to consider rejected papers, as they are not made public.

We convert initial-round reviewer reports (before author revisions) from PDF to structured JSON via RegEx and manual curation. The final dataset comprises 31 papers; additional selection details are provided in \appref{appendix:datasets_choice_of_papers}. Papers contain $11{,}023 \pm 3{,}534$ words and $13.4 \pm 10.0$ figures on average and received $3.1 \pm 1.1$ reviewers.

\subsection{Peer Review Generation}
For both venues, we generate reviews using Llama 4 Scout, Qwen3-VL-235B-Thinking, and GPT-5. The former two are state-of-the-art open-weight multimodal models, whereas GPT-5 serves as a proprietary baseline. We do not enable web search to prevent models from finding original reviews or acceptance decisions. Implementation details and exact model versions are provided in \appref{app:model-specs}.

Following \citet{tyser2024ai}, we incorporate venue-specific reviewer materials, including reviewing guidelines, evaluation criteria, and policy documents, into the prompts for both venues. We use their prompts as a starting point, adapting them to our model suite and extending the generation procedure to reflect each venue's review format. For ICLR, we generate the individual review fields sequentially, conditioning each field on previously generated content to promote consistency. For Nature Communications, we generate a single narrative review report in one pass, reflecting the journal's unstructured review format. Details on manuscript preprocessing, prompts, and supporting materials are provided in \appref{app:review-generation}.

\subsection{Evaluation Metrics}
Because peer-review quality is multi-dimensional, we evaluate generated reviews using multiple complementary metrics that capture different aspects of review quality.

\subsubsection{Score Alignment}
A natural way to assess the alignment between human and LLM-generated peer reviews is to compare their score distributions and correlations with aggregated human scores for the same paper. This analysis is limited to ICLR, as Nature Communications does not include numerical review scores.

\subsubsection{Overlap Score}
Aggregate quality scores do not reveal whether reviews identify the same substantive issues as human reviewers. Inspired by atomic decomposition \citet{min2023factscore} and the review comparison approach of \citet{liang2024can}, we decompose reviews into concerns using GPT-5.5 and classify them as strengths or weaknesses with severity levels. Since strengths are difficult to compare reliably due to broad positive statements, we focus on weakness coverage: the proportion of pooled human weaknesses recovered by a review.

Because AI-generated reviews are typically longer, we additionally report length-normalized coverage. For each paper, we restrict the evaluated review to $K$ weakness concerns, where $K$ is the average number of weaknesses contributed by one human reviewer for that paper, and average coverage across all size-$K$ subsets. Human reviews use the same leave-one-out setting. Further details are provided in \appref{app:overlap}.

\subsubsection{Granuscore}
We evaluate review granularity using Granuscore \citep{granuscore}, which measures the semantic granularity of text. Lower scores indicate finer-grained feedback, whereas higher scores correspond to more abstract statements. This allows us to compare whether reviews provide fine-grained feedback or remain at a more general level.

\subsubsection{LLM-as-a-Judge}
Finally, we evaluate overall review quality using an LLM-as-a-Judge. Since collecting expert preference judgments for every generated review is impractical, we use GPT-5 with a fixed evaluation rubric and schema-constrained output. The judge is given the full manuscript in Markdown form together with a single review. Following \citet{tyser2024ai}, it assigns five-point ratings for understanding, coverage, evidence support, constructiveness, and conciseness, as well as an overall score on a seven-point scale. Implementation details are provided in \appref{app:reviewevaluation}.

\section{Results}\label{section:evaluation}
This section presents the evaluation results for the generated reviews. For readability, figures showing per-paper results display only the first 20 papers.

\subsection{Review Length}
Before evaluating review quality, we compare the lengths of generated and human reviews. \autoref{tab:generated_stats} reports the average review length across models and datasets. All LLMs generate substantially longer reviews than human reviewers, with GPT-5 consistently producing the longest reviews. For all three models, reviews are markedly shorter for Nature Communications than for ICLR, likely reflecting the difference between ICLR's structured review form and the journal's narrative format. Human reviews do not follow this pattern: those for Nature Communications are slightly longer on average than those for ICLR.

\setlength{\belowcaptionskip}{-10pt}
\begin{table}[t]
\centering
\small
\begin{tabular}{
l
S[table-format=4.0(3)]
S[table-format=4.0(3)]
}
\toprule
\textbf{Source} & {\textbf{ICLR 2026}} & {\textbf{Nature Comm.}} \\
\midrule
GPT   & 3447(258) & 1306(124) \\
Llama & 1893(297) &  992(408) \\
Qwen3 & 2579(337) &  913(123) \\
Human &  454(230) &  550(362) \\
\bottomrule
\end{tabular}
\caption{Review length in words (mean $\pm$ std.).}
\label{tab:generated_stats}
\end{table}

\subsection{Score Alignment}
\label{section:score-alignment}
Ideally, the scores assigned by LLMs should align with those of human
reviewers. Because Nature Communications reviews do not
include numerical paper scores, we restrict this analysis to ICLR. Human reviewers assigned a mean score of $4.3 \pm 1.9$, while GPT gave $6.8 \pm 1.0$, and both Llama and Qwen gave $7.9 \pm 0.8$ and $7.9 \pm 0.7$. This is far above the human baseline.

To assess whether the models nevertheless rank papers consistently with human reviewers, we correlate their scores with the mean human score for each paper. GPT shows moderate alignment (Pearson $r=0.62$, Spearman $\rho=0.60$; both $p<0.001$), suggesting that it captures some signal about relative paper quality despite its upward bias. In contrast, Llama ($r=-0.01$, $\rho=-0.02$) and Qwen ($r=0.13$, $\rho=0.09$) show no significant alignment ($p>0.5$). Correlations using median human scores yield the same conclusion. A per-paper visualization of the assigned scores and submission decisions is provided in
\appref{app:results-score-alignment}.

\autoref{tab:given_score_by_status} further compares accepted vs. rejected submission scores. Human reviewers separate the groups by $1.53$ points on average. GPT reproduces roughly half of this gap ($0.87$ points), whereas Llama and Qwen show almost no separation, meaning their scores are not only inflated but also largely uninformative about the acceptance decision.

\setlength{\belowcaptionskip}{-10pt}
\begin{table}[t]
\small
\centering
\begin{tabular}{@{}l S[table-format=1.2(1.2)] S[table-format=1.2(1.2)] S[table-format=+1.2]@{}}
\toprule
\textbf{Source} & {\textbf{Accepted} ($n{=}21$)} & {\textbf{Rejected} ($n{=}29$)} & {$\Delta$} \\
\midrule
GPT   & 7.29(0.64) & 6.41(1.09) & +0.87 \\
Llama & 8.00(0.32) & 8.03(0.19) & -0.03 \\
Qwen  & 7.95(0.38) & 7.86(0.74) & +0.09 \\
Human & 5.18(0.79) & 3.65(1.12) & +1.53 \\
\bottomrule
\end{tabular}
\caption{Assigned scores for accepted and rejected ICLR submissions
(mean $\pm$ std.\ across papers). Human scores are averaged across reviewers
per paper before aggregation. $\Delta$ denotes the mean score for
accepted submissions minus that for rejected submissions.}
\label{tab:given_score_by_status}
\end{table}

\subsection{Overlap Score}
\begin{figure}[t]
    \centering
    \includegraphics[width=\columnwidth]{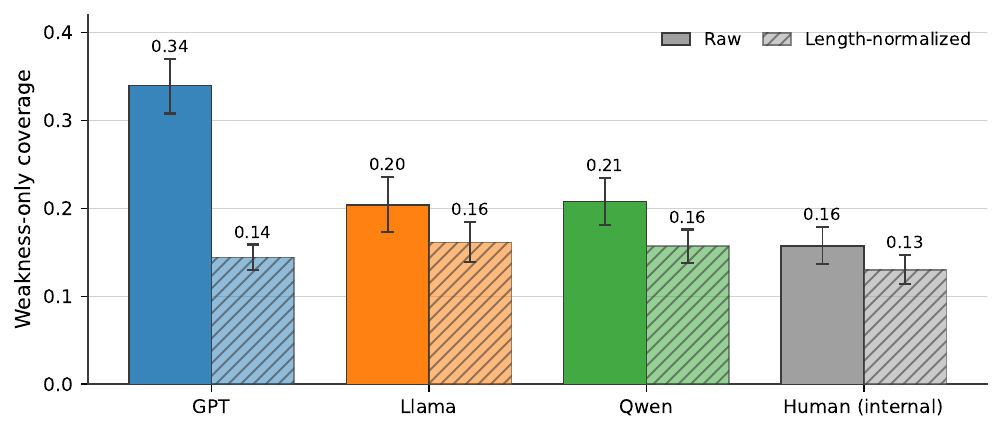}
    \caption{Raw and length-normalized weakness coverage across review sources. Confidence intervals are obtained via a paper-level bootstrap (2,000 resamples).}
    \label{fig:lengthnorm}
\end{figure}

We evaluate how much of the pooled human concerns each review reproduces. \autoref{fig:lengthnorm} summarizes the raw and length-normalized coverage. Raw weakness coverage is highest for GPT, followed by Llama and Qwen, while the human leave-one-out baseline is lower. This difference is partly explained by the substantially larger number of concerns generated by AI reviewers: GPT produces a median of 36 extracted concerns per review compared with 11 for humans. When controlling for this difference in review length, the gap is much smaller: all sources fall into a narrow range, with GPT becoming comparable to the human baseline.

The extracted concerns also reveal a difference in recommendation behavior. Human reviewers provide an explicit recommendation for 69\% of their weaknesses, while AI reviewers almost always attach a proposed action (GPT: 100\%, Llama: 98\%, Qwen: 97\%). AI reviews also differ in the type of actions proposed, with humans more frequently requesting clarifications or textual revisions, whereas AI reviews more often suggest additional experiments or analyses (\appref{app:overlap-recoms}). We also examine the types of human concerns captured. \autoref{tab:results_seriousness} shows that coverage increases with concern seriousness: GPT covers 46\% of Essential concerns, compared with 34\% of Recommended and 12\% of Minor concerns. Llama and Qwen show the same trend, although at lower overall coverage. Finally, overlap varies across reviewing settings: coverage decreases
substantially from ICLR to Nature Communications for all sources, dropping by roughly one third to one half (e.g., GPT: $0.52\rightarrow0.33$ and human baseline: $0.27\rightarrow0.14$; \appref{app:overlap-venue}).

\setlength{\belowcaptionskip}{-12pt}
\begin{table}[t]
\centering
\small
\begin{tabular}{lcccr}
\toprule
\textbf{Seriousness} & \textbf{GPT} & \textbf{Llama} & \textbf{Qwen} & $n$ \\
\midrule
Essential   & 0.46 & 0.22 & 0.26 & 514 \\
Recommended & 0.34 & 0.23 & 0.22 & 1536 \\
Minor       & 0.12 & 0.01 & 0.02 & 382 \\
\bottomrule
\end{tabular}
\caption{Coverage by the seriousness of the human concern; $n$ is the number of human concerns.}
\label{tab:results_seriousness}
\end{table}

\subsection{Granuscore}
\begin{figure*}[t]
    \centering
    \includegraphics[width=1\textwidth]{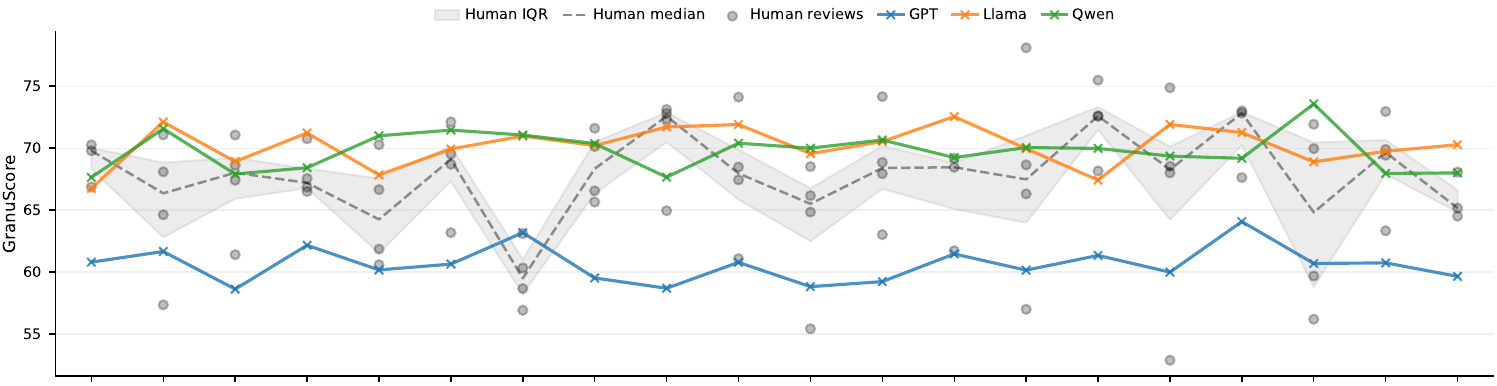}
    \caption{Per-paper GranuScore for ICLR reviews. Gray points show human reviews, with their median and interquartile range indicated by the dashed line and shaded band; colored lines show generated reviews.}
    \label{fig:granuscore_per_paper_ICLR} 
\end{figure*}

\autoref{fig:granuscore_per_paper_ICLR} shows the GranuScore distributions for ICLR reviews. Human reviews score \(68.4 \pm 4.7\), compared with \(60.7 \pm 1.7\) for GPT, \(70.1 \pm 2.2\) for Llama, and \(70.0 \pm 1.7\) for Qwen. The same pattern largely holds for \textit{Nature Communications} (Human: \(65.5 \pm 5.6\); GPT: \(61.5 \pm 2.9\); Llama: \(73.0 \pm 3.3\); Qwen: \(67.9 \pm 3.3\)); the figure is provided in \appref{app:granuscore-results}. Across both venues, Llama and Qwen receive higher GranuScore values than human reviews, whereas GPT receives lower values.

\subsection{LLM-as-a-Judge}

\setlength{\belowcaptionskip}{-15pt}
\begin{figure*}[t]
    \centering
    \includegraphics[width=1\textwidth]{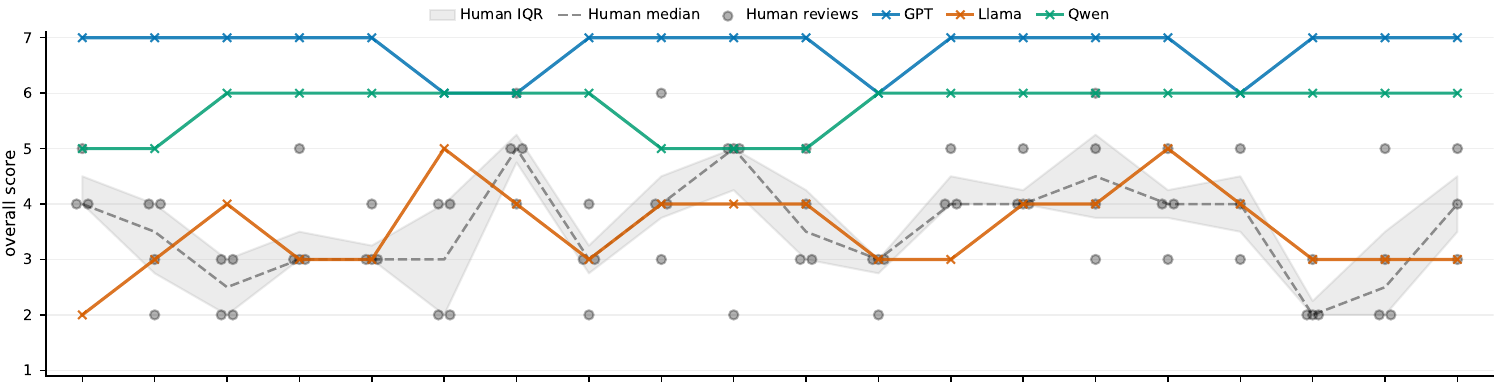}
    \caption{Overall score given by LLM-as-a-Judge for the peer-reviews of ICLR papers.} \label{fig:overall_score_ICLR} 
\end{figure*}

\setlength{\belowcaptionskip}{-15pt}
\begin{table*}[t]
\small
\centering
\begin{tabular}{
ll
S[table-format=1.1(1.1)]
S[table-format=1.1(1.1)]
S[table-format=1.1(1.1)]
S[table-format=1.1(1.1)]
S[table-format=1.1(1.1)]
S[table-format=1.1(1.1)]
}
\toprule
\textbf{Dataset} &
\textbf{Source} &
{\textbf{Overall}} &
{\textbf{Underst.}} &
{\textbf{Coverage}} &
{\textbf{Evid.\ Sup.}} &
{\textbf{Constr.}} &
{\textbf{Concis.}} \\
\midrule
\multirow{4}{*}{ICLR 2026}
 & GPT   & 6.7(0.4) & 5.0(0.0) & 5.0(0.0) & 5.0(0.0) & 5.0(0.0) & 4.0(0.0) \\
 & Llama & 3.5(0.8) & 3.4(0.9) & 3.6(0.6) & 2.4(0.7) & 3.7(0.7) & 3.0(0.7) \\
 & Qwen  & 5.8(0.5) & 4.9(0.3) & 4.6(0.5) & 4.4(0.7) & 5.0(0.1) & 4.0(0.4) \\
 & Human & 3.5(1.1) & 3.2(1.0) & 3.2(0.9) & 2.5(0.9) & 3.5(0.9) & 4.4(0.6) \\
\midrule
\multirow{4}{*}{Nat.\ Comms}
 & GPT   & 6.3(0.4) & 5.0(0.0) & 5.0(0.0) & 5.0(0.0) & 5.0(0.0) & 4.1(0.3) \\
 & Llama & 3.0(0.7) & 3.6(0.6) & 3.0(0.6) & 2.4(0.5) & 2.9(0.5) & 2.3(1.1) \\
 & Qwen  & 5.9(0.3) & 4.9(0.2) & 4.9(0.2) & 4.7(0.4) & 5.0(0.0) & 4.0(0.0) \\
 & Human & 4.6(1.2) & 4.0(1.0) & 3.8(1.0) & 3.5(1.0) & 4.5(0.8) & 4.4(0.8) \\
\bottomrule
\end{tabular}
\caption{LLM-as-a-Judge scores (mean $\pm$ std.\ across papers) by source and dataset. Overall is rated on a 1--7 scale; all other dimensions are rated on a 1--5 scale.}
\label{tab:llmjudge_scores}
\end{table*}

Finally, we evaluate review quality using an LLM judge.
\autoref{tab:llmjudge_scores} reports the aggregate overall and
dimension-level scores, while \autoref{fig:overall_score_ICLR} shows the
per-paper distribution of overall scores for ICLR. The corresponding overall-score distribution for Nature Communications, together with the per-paper distributions for all five dimensions in both venues, is provided in \appref{app:llm-as-a-judge-results}. Across both venues, the judge assigns
substantially higher overall scores to GPT- and Qwen-generated reviews
than to human reviews. Llama matches human performance on ICLR but scores
considerably lower on Nature Communications.

GPT and Qwen also receive near-maximum scores on most individual dimensions.
Conciseness is the only dimension on which human reviews consistently
outperform all three models, whereas Llama performs particularly poorly on
evidential support and conciseness. These patterns are consistent across the
structured ICLR format and the narrative Nature Communications
reviews.

Judged review quality was largely independent of paper outcome. For ICLR,
mean scores for accepted and rejected submissions differed by at most
0.37 points across all dimensions, with most differences being smaller (\appref{app:llm-as-a-judge-results}).

\subsection{Qualitative Analysis}
We complement the quantitative metrics by a qualitative analysis of
generated reviews. We identify systematic differences in the review
of behavior that are not fully captured by aggregate quality scores. Additional examples and detailed analyses are provided in \appref{app:qualitative-analysis}.

\paragraph{Positive Bias}
\autoref{section:score-alignment} already showed that LLMs assign systematically higher ratings than human reviewers. This tendency is also visible in the free-text reviews. Llama frequently closes reviews with overt recommendations for publication, even though the prompt does not require an accept/reject decision. Across ICLR, $47/50$ Llama reviews contain phrases such as ``highly recommend'', ``suitable for publication'', or ``significant contribution'', compared with $16/50$ GPT, $21/50$ Qwen and $44/192$ human reviews. A similar pattern is observed for Nature Communication reviews (\appref{app:pos-bias}).

\paragraph{Evidence Grounding}
Generated reviews also differ in how explicitly they refer to the manuscript.
GPT produces substantially more references to figures, tables, sections, and
equations than human reviewers, whereas Llama rarely points to specific
locations in the manuscript. Detailed counts are provided in
\appref{app:evidence-grounding}.

\section{Discussion}
\label{section:discussion}
\paragraph{Policies Across Research Communities}
Our survey shows that reviewer AI policies are not uniform across publication
communities. AI/NLP venues are generally more permissive toward limited AI
assistance, whereas medical journals more often prohibit reviewer AI use,
reflecting stronger confidentiality concerns. However, the prevalence of
partial policies also suggests that AI assistance is not necessarily treated
as an all-or-nothing decision: many venues distinguish between different uses,
such as language assistance and processing manuscript content. Future policy
discussions may therefore benefit from evaluating specific use cases and their
associated risks rather than treating reviewer AI use as a single category.

\paragraph{AI as Review Support}
% The overlap analysis suggests that the main value of current AI reviewers is not superior judgment, but breadth. AI-generated reviews recover more human concerns than individual reviewers in raw coverage, but this advantage largely disappears when controlling for review length. Thus, LLMs appear to generate a larger set of candidate issues rather than consistently identifying concerns better than human reviewers. However, this increased breadth also introduces uncertainty: many AI-generated concerns are not raised by human reviewers, and their validity requires further verification.

%Importantly, the additional concerns identified by AI systems are not limited to superficial observations, contrasting with earlier findings that LLMs struggle with deeper critique \citep{robertson2023gpt4}. GPT recovers a larger fraction of Essential human concerns than Recommended or Minor concerns. Beyond identifying potential weaknesses, AI systems may also support reviewers in formulating more actionable feedback: AI reviews attach recommendations to almost all identified weaknesses, whereas human reviewers provide explicit recommendations less consistently.

%This suggests a role for AI as a second-pass review assistant. Rather than exposing reviewers to AI feedback before their own assessment, which could reduce independent judgment and reviewer diversity, AI-generated reviews could provide additional candidate concerns after the initial review for reviewers to validate, incorporate, or discard.

The overlap analysis suggests that the main value of current AI reviewers lies
in breadth rather than superior judgment. AI-generated reviews recover more
human concerns than individual reviewers in raw coverage, but this advantage
largely disappears when controlling for review length. LLMs, therefore, expand the set of candidate issues rather than identify the exact same concerns as human reviewers. This breadth may be useful, but it also introduces uncertainty: many generated concerns are not raised by human
reviewers and must be independently verified.

Importantly, these additional concerns are not limited to superficial
observations, contrasting with earlier findings that LLMs struggle with deeper
critique \citep{robertson2023gpt4}. GPT recovers a larger fraction of Essential
human concerns than Recommended or Minor concerns. AI systems may also support
the formulation of more actionable feedback, attaching recommendations
to almost all identified weaknesses, whereas human reviewers provide explicit
recommendations less consistently.

Taken together, these findings support current LLMs as reviewer
assistants rather than autonomous reviewers \citep{sun2025llm}. A suitable
workflow would use AI as a second-pass assistant: reviewers would first form an
independent assessment and only then inspect AI-generated feedback, reducing
anchoring and preserving diversity of judgment. The AI output could serve as
a set of candidate concerns or possible omissions for reviewers to validate,
incorporate, or discard, while scientific judgment and responsibility remain
with the reviewer. This role is consistent with emerging systems such as REVAS
and with interventions that provide feedback on reviewers' own reports rather
than replacing their judgment \citep{thakkar2026large}. Future work should
examine how such workflows affect reviewer effort, feedback quality, review
diversity, and decision-making.

\paragraph{Challenges in Evaluating AI-Generated Reviews}
Our results highlight that AI review quality cannot be captured by a single
metric. LLM-as-a-Judge scores provide an optimistic performance assessment, while score alignment reveals that fluent reviews do not
necessarily correspond to calibrated scientific judgment. In particular,
Llama and Qwen produce highly similar scores across papers, but fail to
distinguish accepted and rejected submissions, whereas GPT shows only moderate
alignment with human assessments. Previous findings
that LLM reviewers tend to produce overly positive evaluations
\citep{latona2024lottery} were also reinforced.

Beyond overall quality scores, review quality also depends on how feedback is
formulated. Granularity and evidence grounding analyses show that reviews
differ in the level at which they express criticism: abstract statements may
appear comprehensive while providing limited guidance, whereas grounded
feedback connects observations to specific manuscript content. Thus, evaluating
AI reviewers requires assessing not only which concerns are identified, but
also how they are communicated.

At the same time, agreement with human reviews should not be considered a
perfect target. Overlap drops substantially from ICLR to Nature Communications
for both AI and human reviewers, and inter-human overlap is itself
limited. Matching human reviewers is therefore an imperfect gold standard and
partly reflects properties of the review setting rather than a single notion
of review quality.

%\paragraph{Towards AI-Assisted Reviewing}
%Taken together, our findings support the view that current LLMs are best considered reviewer assistants rather than autonomous reviewers \citep{sun2025llm}. Their value lies in expanding candidate concerns and supporting feedback formulation, while human reviewers remain responsible for scientific judgment. This aligns with emerging reviewer-assistance systems such as REVAS. Future work should investigate how such human-AI workflows affect reviewer effort, feedback quality, and decision-making.

\paragraph{Takeaways and Benefits for Publishing Venues}
For publishing venues, AI assistance offers a potential way to improve the
scalability and consistency of reviewing as submission volumes continue to
grow \citep{kim2025position}. Its most defensible role is to support labor-intensive parts of the process, such as surfacing overlooked issues, encouraging actionable feedback,
and helping chairs identify areas of agreement and disagreement, instead of automating acceptance decisions. This may allow scarce reviewer attention to remain focused on novelty, significance,
and context-dependent scientific judgment. The AAAI-26 pilot illustrates how such assistance can be introduced while preserving human authority \citep{biswas2026aaai}. 

\section{Conclusion}\label{section:conclusion}
In this work, we studied AI-assisted peer review from two perspectives: (1) We surveyed reviewer AI policies across 111 AI/NLP conferences and medical journals, revealing substantial differences in how communities regulate AI use. (2) We compiled a novel dataset with original paper submissions and reviews and evaluated LLM-review-generation on it. Our results show that current LLMs can produce detailed and highly rated reviews, but exhibit limitations in calibration, grounding, and feedback quality. We find that LLMs provide value by expanding the range of identified concerns and supporting more actionable feedback, but are still not suitable as autonomous reviewers.

\section*{Limitations}\label{section:limitations}

\paragraph{Evaluation Challenges}
Peer review quality is difficult to evaluate because there is no single
correct review for a manuscript. Our metrics capture complementary aspects
of review quality, but each has limitations. For example, concern overlap measures agreement with human reviewers rather than review usefulness itself. On the other hand, LLM-as-a-Judge may favor certain stylistic properties of AI-generated reviews, and their
scores may not fully reflect scientific review quality. This is prevalent especially in our setting, where GPT-5 is used both as a review generator and evaluator, which may
introduce bias toward GPT-like outputs. We counterbalanced this by not relying on LLM-as-a-Judge as our only evaluation. A direct human evaluation by experienced and independent reviewers would further strengthen the suite, but it was too costly for our research team to conduct.

\paragraph{Prompt and Model Configuration Dependence}
The generated reviews depend on the selected models, prompts, and generation
strategies. Although we adapt venue-specific prompts based on prior work,
different prompting approaches or future model versions may produce different
results. Our findings should therefore be interpreted as a comparison of the
evaluated systems rather than an absolute ranking of model capabilities.

\paragraph{Dataset Scope and Generalizability}
Our evaluation is limited to two publication venues: ICLR 2026 and
Nature Communications. While these venues provide complementary
conference and journal settings, they cannot represent the full diversity of
peer-review practices across disciplines, review formats, or research
communities. In particular, our Nature Communications dataset only contains
accepted papers with publicly available transparent reviews, as rejected
submissions are not disclosed. Future evaluations across additional venues and
disciplines are needed to determine how broadly our findings generalize.

\paragraph{Input and Resource Constraints}
Due to computational constraints, models do not receive all information that
may be available during human review. For example, we limit the number of
included figures and do not provide external retrieval. This
may reduce the ability of models to verify citations, assess novelty, or
identify missing related work. However, restricting the input to the
manuscript itself provides a controlled comparison and avoids giving AI
reviewers information unavailable to reviewers at submission time.

\section*{Ethical Considerations}\label{section:ethics_statement}
We recognize that AI-assisted review systems carry substantial risks, including automation bias, overreliance on superficially plausible assessments, and the potential amplification of errors or biases in consequential editorial decisions. Our role in this work is evaluative: we aim to identify the capabilities, limitations, and risks of such systems rather than advocate for their autonomous deployment. Accordingly, the prompting strategies presented in this paper should not be adopted directly in review pipelines that influence acceptance or rejection decisions without additional validation, safeguards, and careful consideration of the surrounding institutional context. As emphasized throughout the manuscript, decision authority and accountability should remain with human reviewers and editors.

\section*{Acknowledgments}\label{section:acknowledgments}
This work used LLM-based tools for language edits, clarity improvements, and coding support. All analysis, research, and ideas are either our own or cited. The research has been funded by the German Federal Ministry of Research, Technology, and Space (BMFTR) through grant 01IS23069, Software Campus 3.0 (Technical University of Munich), as part of the Software Campus project “Know ELViS”, as well as by the public German civil law Barbara-Wengeler-Stiftung.

\bibliography{acl_latex}
\bibliographystyle{acl_natbib}

\clearpage
\onecolumn
\appendix
\section{Model Versions and Inference Configuration}
\label{app:model-specs}

To support reproducibility, \autoref{tab:models} lists the exact model versions and deployment methods used throughout our experiments. The following paragraphs describe the inference configurations for review generation, LLM-as-a-judge evaluation, and the model inside the Overlap Score.

\begin{table*}[ht]%
\small
\centering%
\begin{tabular}{@{}l l l@{}}
\toprule%
Name&Version&Access Provider\\%
\midrule%
Llama 4 Scout~\citep{meta_ai_llama_2025} & Llama-4-Scout-17B-16E-Instruct & local (vllm) \\%
Qwen3-VL-235B~\citep{qwen2025qwen3vl} & Qwen3-VL-235B-A22B-Thinking-FP8 & local (vllm) \\%
GPT-5~\citep{openai2025gpt5} & gpt-5-2025-08-07 & OpenAI API\\%
GPT-5.5~\citep{openai2026gpt55} & gpt-5.5-2026-04-23 & OpenAI API\\%
\bottomrule%
\end{tabular}%
\caption{Specific model versions used in our experiments. For each model we provide the exact version and the access provider.}%
\label{tab:models}
\end{table*}

\paragraph{Review generation.}
We sample Qwen3-VL and Llama 4 Scout with a temperature of $0.6$.
GPT-5 is queried using the OpenAI API defaults: a temperature of $1.0$
and medium reasoning effort, as the model does not support custom
temperature values.

We impose a maximum number of generated tokens for each response.
For ICLR, reviews are assembled from answers generated separately for
each review field. We set the limit to $10{,}000$ tokens for GPT-5,
whose reasoning tokens count toward the output budget, and $2{,}000$
tokens for Qwen3-VL and Llama 4 Scout, whose responses rarely approach
this limit. For Nature Communications, reviews are generated
in a single pass. We therefore use a limit of $10{,}000$ tokens for
the two reasoning models, GPT-5 and Qwen3-VL, and $2{,}000$ tokens for
Llama 4 Scout. The latter limit was selected after manual inspection,
as longer generations consistently degenerated into repetition.

Both open-weight models are served locally using an OpenAI-compatible
vLLM v0.14.0 server, with GPU memory utilization set to $0.90$ and a
maximum model context length of $64{,}000$ tokens. Qwen3-VL is loaded
from its natively FP8-quantized checkpoint and distributed across eight
NVIDIA H100 GPUs using tensor and expert parallelism. Because the model
emits its reasoning trace, we remove this trace from the final review
and store it separately. Llama 4 Scout is served in its native precision
across four NVIDIA H100 GPUs using tensor parallelism.

\paragraph{LLM-as-a-Judge evaluation.}
We constrain the judge output through a forced function call using the
evaluation schema described in \appref{app:reviewevaluation}. We do not
set additional sampling or output-budget parameters; GPT-5 uses the API defaults of temperature $1.0$ and medium reasoning effort,
without an explicit output-token limit.

\paragraph{Overlap Score.}
For concern extraction and matching, we use GPT-5.5 through the OpenAI API. Outputs are constrained using Structured Outputs with a strict JSON schema. We do not set additional sampling or output-budget parameters; GPT-5.5 uses the API defaults of temperature $1.0$ and medium reasoning effort, without an explicit output-token limit.

\section{Selection of Nature Communications Papers}
\label{appendix:datasets_choice_of_papers}

We queried the Research Square API for manuscripts that met three criteria:
(i) they were posted between January 1, 2024, and January 1, 2026;
(ii) they were subsequently published in Nature Communications;
and (iii) their titles contained at least one of the following terms:
\emph{agent}, \emph{alignment}, \emph{BERT}, \emph{few-shot},
\emph{fine-tuning}, \emph{foundation model}, \emph{generative},
\emph{GPT}, \emph{language model}, \emph{pretrained},
\emph{question answering}, \emph{representation learning},
\emph{self-supervised}, \emph{transfer learning}, or
\emph{transformer}. This focus on NLP-related terms reflects the authors' domain expertise
and supports a more reliable manual assessment of review quality.

Each preprint was manually linked to its published DOI. We converted the
manuscript and peer-review PDFs to structured Markdown using Docling
\citep{auer2024doclingtechnicalreport}. From each review file, we extracted only the initial-round
reviewer reports, excluding author responses and subsequent review rounds.
Because the formatting varied across files, all extracted reports were
manually verified and corrected where necessary.

The query returned 33 papers. We excluded two because their publicly
available peer-review records were incomplete, resulting in a final sample
of 31 papers. Of these, 7 were published in 2024 and 24 in 2025.

\section{Review Generation Inputs and Prompts}
\label{app:review-generation}
This section describes manuscript preprocessing, the venue-specific
materials supplied to the models, and the complete prompts used to
generate reviews.

\subsection{Manuscript Preprocessing and Model Input}
Each manuscript is converted to structured Markdown using Docling
\citep{auer2024doclingtechnicalreport}. To reduce inference cost and
context length, the model input includes at most five manuscript figures and excludes selected lower-priority sections. For Nature Communications, we remove reference lists and sections, including acknowledgments, author-contribution statements, competing-interest statements, funding
information, corresponding-author information, additional-information, and data- and code-availability statements, whenever they can be identified from the headings extracted by Docling. These sections are often formulaic and largely repeat publication-related declarations rather than content central to evaluating the paper. They are removed from every manuscript and account for 16\% of the extracted text in aggregate. The manuscripts contain no appendices, as supplementary materials are distributed separately. For 16 ICLR manuscripts, we removed references due to an initially different formatting approach. Appendices remain in the model input because they are part of the submitted manuscript and are likewise available to human reviewers. Within each dataset, all models receive identical manuscript inputs, so differences in preprocessing do not confound comparisons between models. The released dataset additionally contains the unfiltered manuscript text.

\subsection{Prompt Design}
The prompt design builds on \citet{tyser2024ai} and is adapted to the
review format and requirements of each venue. For both datasets, the
prompts incorporate venue-specific reviewer guidelines, evaluation
criteria, and policy documents. During prompt development, we also explored few-shot prompting and a second-pass revision stage. Manual inspection revealed no
clear and consistent benefit, we omitted both approaches from the final
generation pipeline.

\subsection{Venue-Specific Reviewer Materials}
\label{app:usedDocs}
For ICLR, we provide the models with the following materials:
\begin{itemize}
    \item Reviewer Guide: \url{https://iclr.cc/Conferences/2025/ReviewerGuide}
    \item Areachair Guide: \url{https://iclr.cc/Conferences/2025/AreaChairGuide}
    \item Code of Ethics: \url{https://iclr.cc/public/CodeOfEthics}
    \item Code of Conduct: \url{https://iclr.cc/public/CodeOfConduct}
    \item Statistics from ICLR 2024: \url{https://papercopilot.com/statistics/iclr-statistics/}
\end{itemize}
OpenReview data is freely exposed to the public via an open API, the research community broadly treats it as a public domain corpus for non-commercial, academic NLP research.

For Nature Communications, we provide:
\begin{itemize}
    \item Reviewer Instructions: \url{https://www.nature.com/ncomms/for-reviewers/writing-your-report}
    \item Image Standards: \url{https://www.nature.com/ncomms/editorial-policies/image-integrity}
    \item Research Ethics: \url{https://www.nature.com/ncomms/editorial-policies/ethics-and-biosecurity}
    \item Peer Review: \url{https://www.nature.com/ncomms/editorial-policies/peer-review}
\end{itemize}
We have explicitly received permission from Springer Nature to use and share materials.

\subsection{ICLR Prompt Templates}
\label{app:revgenICLR}
\begin{promptboxLONG}[System Prompt: Review Generation for ICLR]
{\fontsize{7}{9}\selectfont
    You are a top-tier academic reviewer.\\ 
    Follow all rules:
    \begin{enumerate}
        \item Use only the provided context. Do not hallucinate.
        \item No chain-of-thought in the final answer.
        \item Write like a human reviewer: Be concise, relevant, specific, and non-repetitive. No boilerplate.
        \item No restating of the questions or instructions.
        \item Cite evidence from context. E.g. the section title or figure id or table id. You can also quote short sentence parts.
        \item For rating/scoring questions (Soundness, Presentation, Contribution, Confidence, Overall): Reason briefly, then give the numeric rating in the form of 'Rating: <value>'.
        \item For Strengths/Weaknesses/Suggestions prefer numbered lists (1., 2., …). Keep the amount of points reasonable. Only keep the least generic points.
        \item You can incorporate Markdown and Latex into your review.
        \item A total review has typically at least 300 and maximal 2000 words. Prefer your reviews to have between 500 and 1500 words.
    \end{enumerate}

    \lbrack Reviewer Guide\rbrack \\
    
    \lbrack Area Chair Guide\rbrack \\
    
    \lbrack Code of Ethics\rbrack \\
    
    \lbrack Code of Conduct\rbrack \\
    
    \lbrack Previous Year Conference Statistics\rbrack
}
\end{promptboxLONG}\label{prompt:systemprompt_iclr}

\begin{promptboxLONG}[Review Field Prompts: Review Generation for ICLR]
{\fontsize{7}{9}\selectfont
    \textbf{Q1 (Summary):} "Briefly summarize the paper and its contributions. Do not critique the paper here. The authors should agree with a well-written summary. Typical range: 3-8 sentences."\\
    
    \textbf{Q2 (Soundness):} "Please assign the paper a numerical rating on the following scale to indicate the soundness of the technical claims, experimental and research methodology and on whether the central claims of the paper are adequately supported with evidence: 4 excellent, 3 good, 2 fair, 1 poor."\\
    
    \textbf{Q3 (Presentation):} "Please assign the paper a numerical rating on the following scale to indicate the quality of the presentation. This should take into account the writing style and clarity, as well as contextualization relative to prior work: 4 excellent, 3 good, 2 fair, 1 poor."\\
    
    \textbf{Q4 (Contribution):} "Please assign the paper a numerical rating on the following scale to indicate the quality of the overall contribution this paper makes to the research area being studied. Are the questions being asked important? Does the paper bring a significant originality of ideas and/or execution? Are the results valuable to share with the broader NeurIPS community? 4 excellent, 3 good, 2 fair, 1 poor."\\
    
    \textbf{Q5 (Strengths):} "Please give a substantive assessment of the strengths of the paper. Be specific, avoid generic remarks. This can, for example, include quality, clarity (e.g. well written), or significance, originality (e.g. a new definition or problem formulation, creative combinations of existing ideas, application to a new domain, or removing limitations from prior results.) If you have many, only list the least generic ones."\\
    
    \textbf{Q6 (Weaknesses):} "Please give a substantive assessment of the weaknesses of the paper. Focus on constructive and actionable insights on how the work could improve towards its stated goals. Be specific, avoid generic remarks. For example, if you believe the contribution lacks novelty, provide references and an explanation as evidence; if you believe experiments are insufficient, explain why and exactly what is missing, etc. Unclarities are adressed later, so only list real weaknesses. If you have many, only list the least generic ones."\\
    
    \textbf{Q7 (Suggestions):} "Please list and carefully describe any questions and suggestions for the authors. Think of the things where a response from the author can change your opinion, clarify a confusion or address a limitation. This is important for a productive rebuttal and discussion phase with the authors. Do not simply repeat all the weaknesses."\\
    
    \textbf{Q8 (Ethics):} "If there are ethical issues with this paper, please flag the paper for an ethics review and select area of expertise that the ethics reviewer should to have. Choose one of the following options: No ethics review needed, Ethics review needed: Discrimination / Bias / Fairness Concerns, Ethics review needed: Inadequate Data and Algorithm Evaluation, Ethics review needed: Inappropriate Potential Applications \& Impact  (e.g., human rights concerns), Ethics review needed: Privacy and Security (e.g., consent, surveillance, data storage concern), Ethics review needed: Compliance (e.g., GDPR, copyright, license, terms of use), Ethics review needed: Research Integrity Issues (e.g., plagiarism), Ethics review needed: Responsible Research Practice (e.g., IRB, documentation, research ethics), Ethics review needed: Failure to comply with NeurIPS Code of Ethics (lack of required documentation, safeguards, disclosure, licenses, legal compliance)"\\
    
    \textbf{Q9 (Overall):} "Please provide an 'overall score' for the paper: 10: Award quality: Technically flawless paper with groundbreaking impact, with exceptionally strong evaluation, reproducibility, and resources, and no unaddressed ethical considerations. 9: Very Strong Accept: Technically flawless paper with groundbreaking impact on at least one area of AI/ML and excellent impact on multiple areas of AI/ML, with flawless evaluation, resources, and reproducibility, and no unaddressed ethical considerations. 8: Strong Accept: Technically strong paper, with novel ideas, excellent impact on at least one area, or high-to-excellent impact on multiple areas, with excellent evaluation, resources, and reproducibility, and no unaddressed ethical considerations. 7: Accept: Technically solid paper, with high impact on at least one sub-area, or moderate-to-high impact on more than one areas, with good-to-excellent evaluation, resources, reproducibility, and no unaddressed ethical considerations. 6: Weak Accept: Technically solid, moderate-to-high impact paper, with no major concerns with respect to evaluation, resources, reproducibility, ethical considerations. 5: Borderline accept: Technically solid paper where reasons to accept outweigh reasons to reject, e.g., limited evaluation. Please use sparingly. 4: Borderline reject: Technically solid paper where reasons to reject, e.g., limited evaluation, outweigh reasons to accept, e.g., good evaluation. Please use sparingly. 3: Reject: For instance, a paper with technical flaws, weak evaluation, inadequate reproducibility and incompletely addressed ethical considerations. 2: Strong Reject: For instance, a paper with major technical flaws, and/or poor evaluation, limited impact, poor reproducibility and mostly unaddressed ethical considerations. 1: Very Strong Reject: For instance, a paper with trivial results or unaddressed ethical considerations."\\
    
    \textbf{Q10 (Confidence):} "Please provide a 'confidence score' for your assessment of this submission to indicate how confident you are in your evaluation. Give at most two brief sentences of justification, then exactly: 'Rating: <1–5>' with the following grounding: 5: You are absolutely certain about your assessment. You are very familiar with the related work and checked the math/other details carefully. 4: You are confident in your assessment, but not absolutely certain. It is unlikely, but not impossible, that you did not understand some parts of the submission or that you are unfamiliar with some pieces of related work. 3: You are fairly confident in your assessment. It is possible that you did not understand some parts of the submission or that you are unfamiliar with some pieces of related work. Math/other details were not carefully checked. 2: You are willing to defend your assessment, but it is quite likely that you did not understand the central parts of the submission or that you are unfamiliar with some pieces of related work. Math/other details were not carefully checked. 1: Your assessment is an educated guess. The submission is not in your area or the submission was difficult to understand. Math/other details were not carefully checked."\\
    
    \textbf{Q11 (Conduct):} "If there are no violations of the Code of Conduct with this paper, please respond with NO. Otherwise, if this paper violates the Code of Conduct, please indicate the relevant section(s) from the following options: 'Yes, Harassment, bullying, or discrimination based on personal characteristics', 'Yes, Inappropriate physical contact, sexual harassment, or unwelcome sexual attention', 'Yes, Offensive comments related to gender, race, religion, or other protected characteristics', 'Yes, Disruption of talks or other events, or behavior interfering with participation', 'Yes, Inappropriate use of imagery, language, or personal attacks in virtual interactions")
}
\end{promptboxLONG}\label{prompt:review-field-iclr}

\subsection{Nature Communications Prompt Templates}
\label{app:prompt:reviewgenNature}
\begin{promptboxLONG}[System Prompt: Review Generation for Nature Communications]
{\fontsize{7}{9}\selectfont
    You are a top-tier academic reviewer.\\
    
    Follow all rules:
    \begin{enumerate}
        \item Use only the provided context. Do not hallucinate.
        \item No chain-of-thought in the final answer.
        \item Write in natural flowing prose like a human expert reviewer. Do not use section headers or a rigid structure. Weave validity, significance, methodology, and suggestions naturally into your text. Be concise, relevant, specific, and non-repetitive. No boilerplate. Not unnecessarily long.
        \item No restating of the questions or instructions.
        \item Cite evidence from context. E.g. the section title or figure id or table id. You can also quote short sentence parts.
        \item For listing strengths/weaknesses/suggestions prefer numbered lists (1., 2., …). Keep the amount of points reasonable. Only keep the least generic points.
        \item You can incorporate Markdown and Latex into your review.
        \item A total review has typically at least 300 and maximal 2000 words. Prefer your reviews to have between 500 and 1500 words.
    \end{enumerate}

    \lbrack image standards\rbrack  \\
    
    \lbrack research ethics\rbrack  \\
    
    \lbrack reviewer report guidelines\rbrack  \\
}
\end{promptboxLONG}\label{PromptSystem:ReviewGen}

\begin{promptboxLONG}[User Prompt: Review Generation for Nature Communications]
{\fontsize{7}{9}\selectfont

\lbrack paper text and referenced images\rbrack \\

\textbf{Criteria for publication}\\
Your review is vital in helping our editors decide if the manuscript meets the journal's criteria for publication, and we ask you to keep the following factors in mind when you write your report:
\begin{itemize}
    \item The quality of the data — whether they are technically sound, obtained with appropriate techniques, analysed and interpreted carefully, and presented in sufficient detail.
    \item The level of support for the conclusions — whether sufficiently strong evidence is provided for the authors' claims and all appropriate controls have been included.
    \item The potential significance of the results — whether these results will be important to the field and advance understanding in a way that will move the field forward. (Note that posting of preprints and/or conference proceedings does not compromise novelty.)
\end{itemize}

The primary purpose of your review is to provide feedback on the soundness of the research reported. This will help authors to improve their manuscript and editors to reach a decision. We do not ask that you make a recommendation regarding publication, but you can set out the arguments for and against publication if you so wish.\\

\textbf{Elements of a reviewer report}\\
In your report, please comment on the following aspects of the manuscript.
\begin{itemize}
    \item Key results: Your overview of the key messages of the study, in your own words, highlighting what you find significant or notable. Usually, this can be summarized in a short paragraph.
    \item Validity: Your evaluation of the validity and robustness of the data interpretation and conclusions. If you feel there are flaws that prohibit the manuscript's publication, please describe them in detail.
    \item Significance: Your view on the potential significance of the conclusions for the field and related fields. If you think that other findings in the published literature compromise the manuscript's significance, please provide relevant references.
    \item Data and methodology: Your assessment of the validity of the approach, the quality of the data, and the quality of presentation. We ask reviewers to assess all data, including those provided as supplementary information. If any aspect of the data is outside the scope of your expertise, please note this in your report or in the comments to the editor. We may, on a case-by-case basis, ask reviewers to check code provided by the authors (see this Nature editorial for more information).
    \item Analytical approach: Your assessment of the strength of the analytical approach, including the validity and comprehensiveness of any statistical tests. If any aspect of the analytical approach is outside the scope of your expertise, please note this in your report or in the comments to the editor.
    \item Suggested improvements: Your suggestions for additional experiments or data that could help strengthen the work and make it suitable for publication in the journal. Suggestions should be limited to the present scope of the manuscript; that is, they should only include what can be reasonably addressed in a revision and exclude what would significantly change the scope of the work. The editor will assess all the suggestions received and provide additional guidance to the authors.
    \item Clarity and context: Your view on the clarity and accessibility of the text, and whether the results have been provided with sufficient context and consideration of previous work. Note that we are not asking for you to comment on language issues such as spelling or grammatical mistakes.
    \item References: Your view on whether the manuscript references previous literature appropriately.
    \item Your expertise: Please indicate any particular part of the manuscript, data or analyses that you feel is outside the scope of your expertise, or that you were unable to assess fully.
\end{itemize}

\textbf{Providing constructive feedback}\\
We ask reviewers to approach peer review with a sincere intention to help the authors improve their manuscripts. Nearly all submissions have weaknesses to be addressed: the best and most constructive reports suggest specific improvements; such feedback can be used by authors to improve their manuscript to the point where it might be suitable for acceptance. Even in instances where manuscripts are rejected, your report will help authors interpret the editor's decision and improve their work prior to submission elsewhere.\\
You should be direct in your report, but you should also maintain a respectful tone. As a matter of policy, we do not censor the content of reviewer reports; any comments that were intended for the authors are transmitted, regardless of what we may think of the content. On rare occasions, we may edit a report to remove offensive language or comments that reveal confidential information about other matters.\\

Now write the full reviewer report in natural flowing prose, as a human expert would. Do not use labeled section headers. Mentally consider all the elements above (validity, significance, methodology, suggested improvements, etc.) but weave them naturally into your writing rather than addressing each as a separate heading.
}
\end{promptboxLONG}\label{PromptUser:ReviewGen}

\section{Review Evaluation}
\label{app:reviewevaluation}
In addition to the numerical ratings, the LLM-as-a-Judge produces textual rationales, a confidence level, and reviewer-oriented feedback. We use the following system and user prompt for the evaluation:

\begin{promptboxLONG}[System Instruction Prompt: LLM-as-a-Judge]
{\fontsize{7}{9}\selectfont
You will be given a PAPER and a REVIEW of that paper.
Rate the QUALITY OF THE REVIEW only (do NOT rate the paper). 
Do not be generic. Be review-specific and precise. Be critical if appropriate. 
Discriminate clearly between strong and weak reviews according to human preference. Avoid clustering most scores around the middle.\\

Agree or disagree with the following statements (5-item Likert):\\

\textbf{Understanding:} The review demonstrates an adequate understanding of the paper.
\begin{itemize}
    \item Review makes comments that are detailed and specific to the paper
    \item It is OK for the reviewer to lack expertise in certain aspects of the paper as long as it is explicitly or implicitly indicated in the review
\end{itemize}

\textbf{Coverage:} The review covers all the required aspects.
\begin{itemize}
    \item Review adequately comments on Soundness, Presentation, and Contribution of the paper
    \item Review adequately comments on Strengths and Weaknesses of the paper
\end{itemize}

\textbf{Evidence Support:} Evaluations made in the review are well supported.
\begin{itemize}
    \item Objective arguments are grounded in the paper's content (e.g., specific results/comparisons) and are correct
    \item Subjective arguments are accompanied with reasoning
    \item A review that brings additional useful information (e.g., counter examples or uncited references which do a part of the claimed work) is especially strong
\end{itemize}

\textbf{Constructiveness:} The review provides constructive feedback to authors.
\begin{itemize}
    \item Whenever possible, critical comments (especially subjective) are accompanied with actionable items on how to improve the paper
    \item Review is unbiased and written in a polite manner
\end{itemize}

\textbf{Conciseness:} The review communicates efficiently without sacrificing substance.
\begin{itemize}
    \item Clear, direct writing with minimal repetition or digressions
    \item Focuses on paper-relevant points; avoids unnecessary verbosity
\end{itemize}

Provide an \textbf{overall score} for the quality of the review:
\begin{itemize}
    \item \textbf{1 = Very low}\\E.g., a generic review that is applicable to any paper / a short and dismissive review.
    \item \textbf{2 = Low}\\E.g., a review with serious flaws on multiple aspects.
    \item \textbf{3 = Fair}\\E.g., a review with serious flaws on one aspect / a review without serious flaws but with limited insights for Authors/Area Chairs.
    \item \textbf{4 = Good}\\E.g., acceptable review, but nothing stands out. Moderately helpful for decision-making.
    \item \textbf{5 = Very good}\\E.g., a helpful review that stands out on some aspects and provides useful insights.
    \item \textbf{6 = Excellent}\\E.g., a very insightful review that stands out on all aspects.
    \item \textbf{7 = Exceptional}\\E.g., an excellent review that helps authors to non-trivially improve the paper / brings a unique piece of information that is crucial for the decision.
\end{itemize}

Also produce an \textbf{'open\_feedback'} field: 1-5 sentences of free-form feedback to reviewers about the review's quality and how to improve it.
}
\end{promptboxLONG}\label{Prompt:LLM-as-a-Judge}

\section{Overlap Score}
\label{app:overlap}
This section provides additional implementation details for the
overlap metric. We first describe the LLM-based concern extraction and matching pipeline, including the prompts used in our experiments. We then examine the sensitivity of weakness coverage to alternative concern-matching methods.

\subsection{Concern Extraction and Matching}
\label{app:overlap-prompts}
The pipeline consists of two stages. Stage~1 decomposes each review into self-contained items and classifies each by channel, type, recommendation, and seriousness. Stage~2 matches the concerns extracted from a generated review against the pooled human concerns for the same paper. We use the same matcher for the human leave-one-out baseline, treating the held-out human review as the ``AI'' concern list and the remaining human reviews as the reference set.

During development, we manually inspected extraction and matching outputs
for outlier papers and iteratively refined the prompts until the inspected
outputs aligned with our judgments.

We show the system prompts below. The output schema is appended programmatically and omitted for brevity. The complete implementation is available in \texttt{extract\_concerns.py} and \texttt{match\_concerns.py}. The channel termed \emph{weakness} in the main text is labeled \texttt{critique} in the prompts; both denote the same.

\begin{promptboxLONG}[System Instruction Prompt: Concern Extraction]
{\fontsize{7}{9}\selectfont
You are an expert analyst of academic peer reviews. Your task is to decompose a single peer review into a structured list of individual concerns and recommendations and classify each one.\\

INPUT\\
You will be given the text of one peer review of an academic paper. The review may be structured (with sections like Summary, Strengths, Weaknesses, Questions) or free-form prose. The paper itself is NOT provided. You are decomposing what the reviewer wrote, not fact-checking it against the paper.\\

OUTPUT\\
Reply with a JSON object that matches the provided schema. The object has a single field "items" containing a list of decomposed items.\\

CHANNEL CLASSIFICATION (every item gets exactly one)
\begin{itemize}
    \item "critique": the reviewer identifies a problem, weakness, missing piece, error, methodological doubt, or thing the author should fix. Includes questions that imply a gap or a concern (e.g., ``How does this handle X?'' when X is a plausible failure mode).
    \item "strength": the reviewer praises something the paper does well (clear writing, strong results, interesting idea, sound methodology, etc.).
    \item "neutral": neither praise nor critique. Use this for a descriptive or factual observation without judgement that appears outside the summary (for example, a neutral restatement embedded in Weaknesses or Questions, or an informational question that does not imply a gap). Do NOT emit neutral items for the review's Summary section, which is skipped entirely (see the summary rule below). Editorial score lines like ``Soundness: 4'' are skipped entirely.
\end{itemize}

DECOMPOSITION RULES
\begin{enumerate}
    \item SKIP THE SUMMARY ENTIRELY. Do not emit any item that comes from the review's summary. For structured reviews, this is the field labelled ``Summary''; for narrative journal reviews, it is an opening paragraph that merely restates or paraphrases the paper without evaluation. By venue guidelines, a summary should be description only, so it contributes no concerns, recommendations, or strengths. Exception: if a summary paragraph contains an explicit evaluative judgement (praise or critique), emit only that evaluative point and ignore the descriptive remainder.
    \item Each item makes EXACTLY ONE point. Do not bundle two distinct concerns or two distinct strengths into one item.
    \item Each item must be understandable on its own. Spell out implicit references (e.g., replace ``this issue'' with the specific issue).
    \item Extract only what the reviewer actually wrote. Do not invent points the reviewer did not raise. Do not soften or sharpen the framing.
    \item If a single sentence contains multiple distinct points, split it into multiple items.
    \item If multiple sentences make the same point with different wording, merge them into one item with a single combined formulation.
    \item Skip boilerplate: greetings, signoffs, conference-system scaffolding, numerical score lines (e.g., ``Soundness: 4'', ``Confidence: 3'', ``Rating: 6'').
    \item Keep the item text concise (one to three sentences). The verbatim quote is not required; a faithful paraphrase is fine as long as the point is preserved.
\end{enumerate}

TYPE CLASSIFICATION (every item gets exactly one)
\begin{itemize}
    \item Experimental design: experiments, baselines, ablations, controls, hyperparameter sweeps, evaluation setup
    \item Methodology \& soundness: correctness of the proposed method, theorems, proofs, claims, logical structure, mathematical rigor
    \item Data quality \& validation: dataset choice, splits, biases, leakage, label noise, sample size
    \item Reproducibility: code or data availability, hyperparameter reporting, deterministic seeding, environment details
    \item Related work: coverage of cited work, accuracy of citations, missing references, mischaracterization of prior work
    \item Clarity \& writing: readability, structure, notation, figure quality, English/grammar
    \item Significance \& contribution: novelty, impact, scope, relevance to the venue, generality of the contribution
    \item Ethics \& broader impact: ethical concerns, dual-use risk, fairness, environmental cost, societal harm
    \item Other: anything that does not fit a category above (use sparingly)
\end{itemize}

RECOMMENDATION FIELDS (only for channel="critique"; set to null for strength and neutral)
\begin{itemize}
    \item recommendation (free text or null): the concrete action the reviewer EXPLICITLY recommends the authors take. Use the reviewer's stated action, paraphrased only for self-containment, in one or two sentences. If the reviewer flagged a problem without proposing an action, set this to null. Do NOT infer or invent recommendations; do not fill this field just because a fix seems obvious.
    \item recommendation\_category (one of the listed options, or null): pick a category only when recommendation is non-null. If \texttt{recommendation} is null, set this to null as well. Categories:
    \begin{itemize}
        \item New experiment: run a new experiment or ablation
        \item New analysis: re-analyse existing results, run additional statistical tests
        \item Textual revision: rewrite, restructure, or improve the prose of a section
        \item Clarification: explain or define something more clearly (notation, claim, motivation)
        \item Additional citations: cite missing related work
        \item Methodology change: alter the proposed method itself
        \item Data: collect more data, change the dataset, fix labels
        \item None: only when the reviewer explicitly recommends something but it does not fit any category above (very rare)
    \end{itemize}
\end{itemize}

SERIOUSNESS FIELDS (only for channel="critique"; set to null for strength and neutral)
\begin{itemize}
    \item seriousness\_explanation (free text, one or two sentences): WHY the critique matters at the seriousness level you assign.
    \item seriousness\_category (one of three):
    \begin{itemize}
        \item Essential: the critique must be addressed for the paper's stated conclusions to hold or for acceptance to be justified. Without addressing it, a major claim of the paper is unsupported.
        \item Recommended: addressing the critique would substantially improve the paper, but the core contribution survives if it is not addressed.
        \item Minor: small improvement; the paper is fine without addressing it.
    \end{itemize}
\end{itemize}

NUMBERING\\
Number items 1, 2, 3, \ldots{} in the order they appear in the review text.\\

GENERAL GUIDANCE
\begin{itemize}
    \item Err on the side of more items rather than fewer, but do not split a single coherent point into trivial fragments.
    \item Be precise about channel: a sentence that praises one thing whilst critiquing another should yield two items, one strength and one critique.
    \item If the review is very short or contains only a numerical score and no text, return an empty items list.
\end{itemize}
}
\end{promptboxLONG}
\label{prompt:concern-extraction}

\begin{promptboxLONG}[System Instruction Prompt: Concern Matching]
{\fontsize{7}{9}\selectfont
You are an expert meta-reviewer. You are given the full text of an academic paper, a list of HUMAN review concerns about that paper, and a list of AI review concerns about the same paper. Each concern has a number, a channel (critique or strength), a type, and a short text.\\

TASK\\
Identify every pair of an AI concern and a human concern that expresses the SAME underlying point about the paper, even when the concerns are worded differently or stated at different levels of granularity. Use the paper text to disambiguate: differently phrased concerns count as the same point only when they refer to the same aspect of this specific paper.\\

RULES
\begin{enumerate}
    \item POLARITY: Match a critique only to a critique and a strength only to a strength. NEVER pair a critique with a strength.
    
    \item SAME POINT: A pair is valid when both concerns address the same substantive issue, gap, claim, contribution, or strength of the paper. Vague thematic overlap (e.g., both concerns mention ``experiments'') is NOT sufficient. They must express the same specific point.
    
    \item GRANULARITY: Matches may be many-to-many. One AI concern may match several human concerns (e.g., when an AI concern combines two points raised separately by human reviewers), and one human concern may match several AI concerns. Emit each valid pair separately.
    
    \item CONFIDENCE: Emit a pair only when you are reasonably confident that the concerns express the same point. When in doubt, do not emit the pair. Unmatched concerns do not appear in the output.
    
    \item Provide a one-sentence justification for each pair that identifies the shared point.
    
    \item If no concerns match, return an empty list.
\end{enumerate}

OUTPUT\\
Reply with a JSON object that matches the provided schema. The object has a single field, "matches", containing a list of objects with the fields "ai\_number", "human\_number", and "justification". The value of "ai\_number" must be a number from the AI concern list, and the value of "human\_number" must be a number from the human concern list.
}
\end{promptboxLONG}
\label{prompt:concern-matching}

\subsection{Choice of Weakness-Only Coverage}
\label{app:overlap-weakness-only}

Although we initially considered both strengths and weaknesses, we use
weakness-only coverage as the primary overlap metric. Strengths are less
suitable for this comparison because broad positive statements can match many
specific human observations. For example, on paper \texttt{OWHKdYwYiF}, a
single broad Llama-4-Scout-17B strength was matched to six separate human
strength items, allowing one general statement of praise to receive credit for
multiple specific observations.

This effect is substantially weaker for weaknesses, where matched concerns
typically correspond to a specific limitation or missing element identified
by reviewers. Furthermore, the composition of review content differs across
sources: human reviews contain 2.3 weaknesses per strength, whereas GPT,
Llama and Qwen produce ratios of 1.6, 0.6, and 0.8,
respectively.

\FloatBarrier
\subsection{Recommendation Behaviour}
\label{app:overlap-recoms}

The concern extraction pipeline records whether a reviewer
explicitly proposes an action for a raised weakness and, if so, the type of
recommended action. \autoref{tab:results_recommendations} reports the resulting recommendation behavior across sources.

\begin{table}[ht]
\centering
\small
\begin{tabular}{lcccc}
\toprule
 & \textbf{Human} & \textbf{GPT} & \textbf{Llama} & \textbf{Qwen} \\
\midrule
Weaknesses with explicit rec. & 69\% & 100\% & 98\% & 97\% \\
\midrule
New experiment       & 21\% & 31\% & 30\% & 30\% \\
New analysis         & 15\% & 30\% & 22\% & 31\% \\
Clarification        & 40\% & 28\% & 28\% & 30\% \\
Textual revision     & 16\% & 5\%  & 6\%  & 4\%  \\
Methodology change   & 2\%  & 3\%  & 10\% & 4\%  \\
Additional citations & 4\%  & 0\%  & 1\%  & 1\%  \\
Data                 & 1\%  & 1\%  & 2\%  & 0\%  \\
\bottomrule
\end{tabular}
\caption{Recommendation behaviour by review source. The first row shows the
fraction of extracted weaknesses accompanied by an explicit recommendation.
The remaining rows show the distribution of recommendation categories among
weaknesses with an explicit recommendation. Columns sum to 100\% up to
rounding; the rare fallback category \textit{None} is omitted.}
\label{tab:results_recommendations}
\end{table}

\subsection{Coverage by Venue}
\label{app:overlap-venue}

\begin{figure}[ht]
    \centering
    \includegraphics[width=0.55\textwidth]{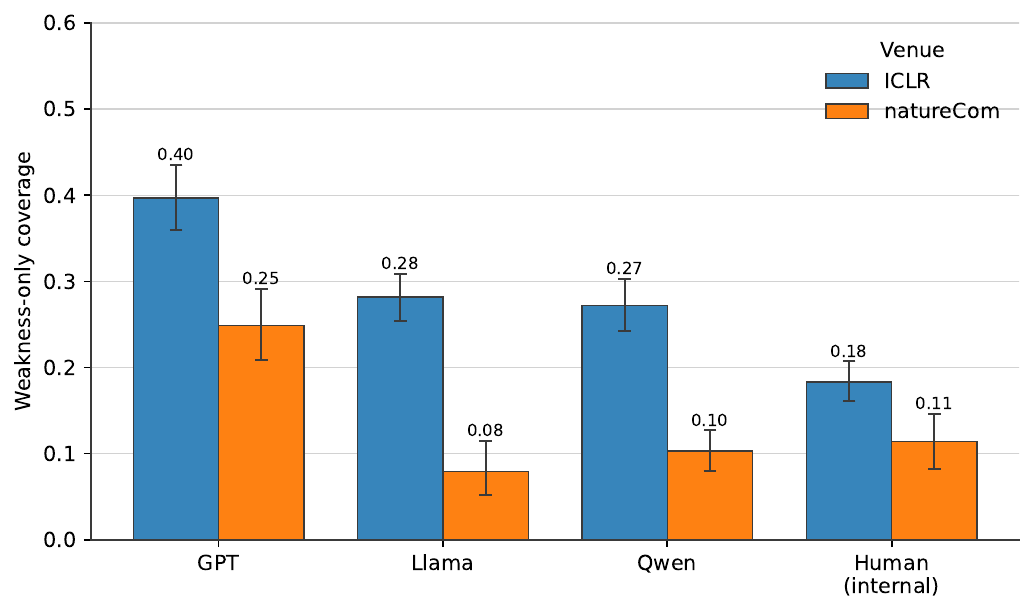}
    \caption{Weakness coverage by venue and source.}
    \label{fig:coverage_by_venue}
\end{figure}

\autoref{fig:coverage_by_venue} shows weakness coverage across the two
reviewing venues. Coverage is consistently lower for Nature Communications
than for ICLR across all review sources, including the human leave-one-out baseline.

\subsection{Comparison of Matching Methods}
\label{app:matching-methods}
To assess how sensitive weakness coverage is to the choice of concern
matcher, we recompute it from the same extracted concerns using two less
costly alternatives: token-level Jaccard similarity and cosine similarity
between sentence embeddings from \href{https://huggingface.co/sentence-transformers/all-MiniLM-L6-v2}{all-MiniLM-L6-v2} via SentenceTransformers \citep{reimers-2019-sentence-bert}.

\begin{figure}[ht]
    \centering
    \includegraphics[width=0.55\textwidth]{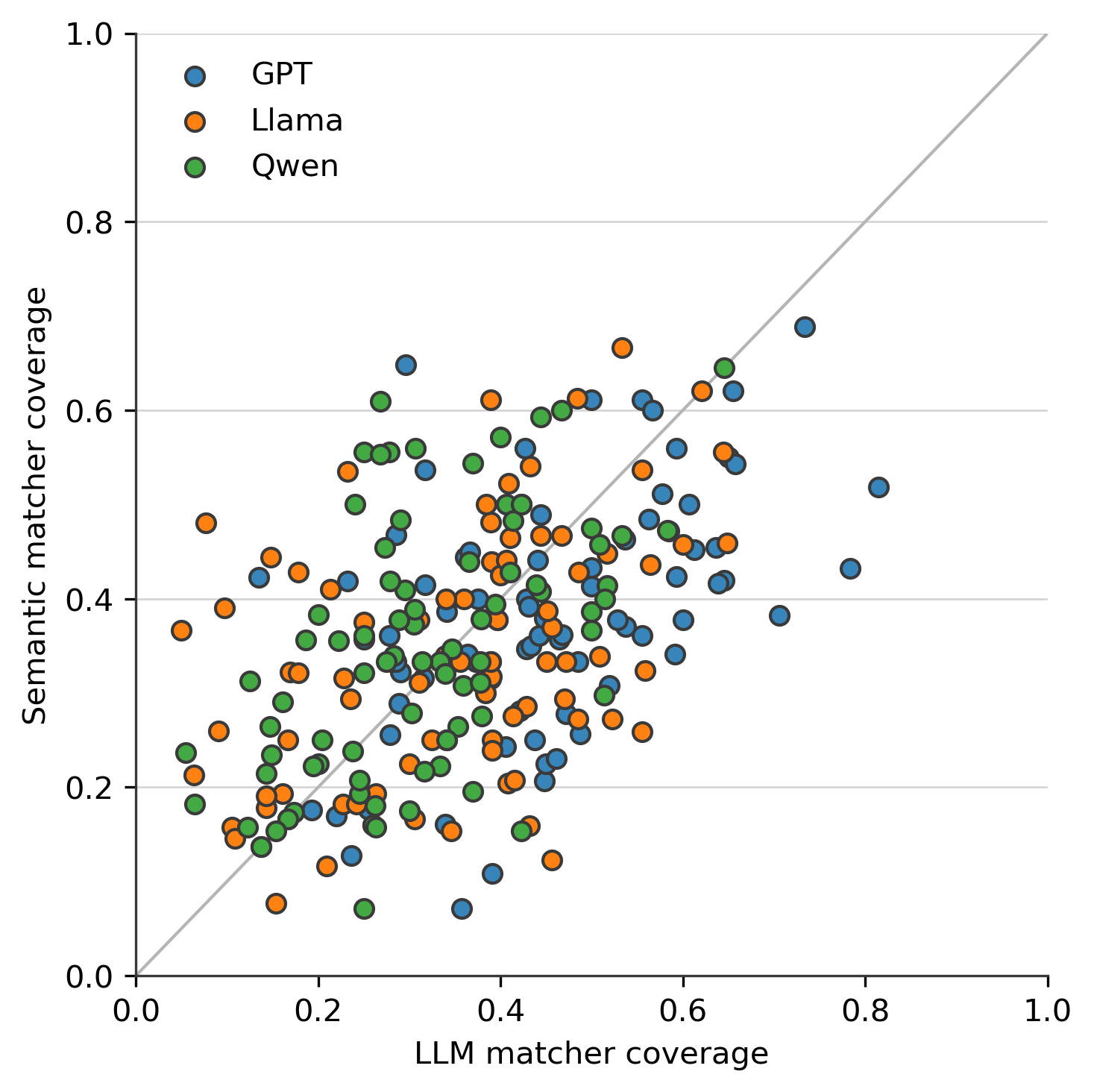}
    \caption{Per-paper weakness coverage obtained with the LLM
    matcher (x-axis) and the sentence-embedding matcher (y-axis). Although the two methods
    produce similar aggregate coverage, their agreement varies across indiviual papers}
    \label{fig:llm_vs_embedding}
\end{figure}

The embedding matcher closely reproduces the aggregate coverage obtained
with the paper-aware LLM matcher ($0.36$ vs.\ $0.37$), suggesting that
the headline coverage estimate is robust to the matching method. Agreement
is lower, however, at the individual-paper and concern-pair levels
(per-paper correlation: $0.50$; pairwise set agreement: $0.39$;
\autoref{fig:llm_vs_embedding}). The lexical matcher yields substantially
lower coverage ($0.13$), indicating that surface-level token overlap
captures only a limited share of the semantic overlap between concerns.

Embedding similarity therefore provides a reasonable approximation of
aggregate weakness coverage but does not recover the same individual
matches. We retain the paper-aware LLM matcher for the main analysis,
particularly for the seriousness- and concern-type-specific breakdowns,
where access to the manuscript can help disambiguate related concerns.

\section{Detailed ICLR Score Alignment Results}
\label{app:results-score-alignment}
\begin{figure*}[ht]
    \centering
    \includegraphics[width=1\textwidth]{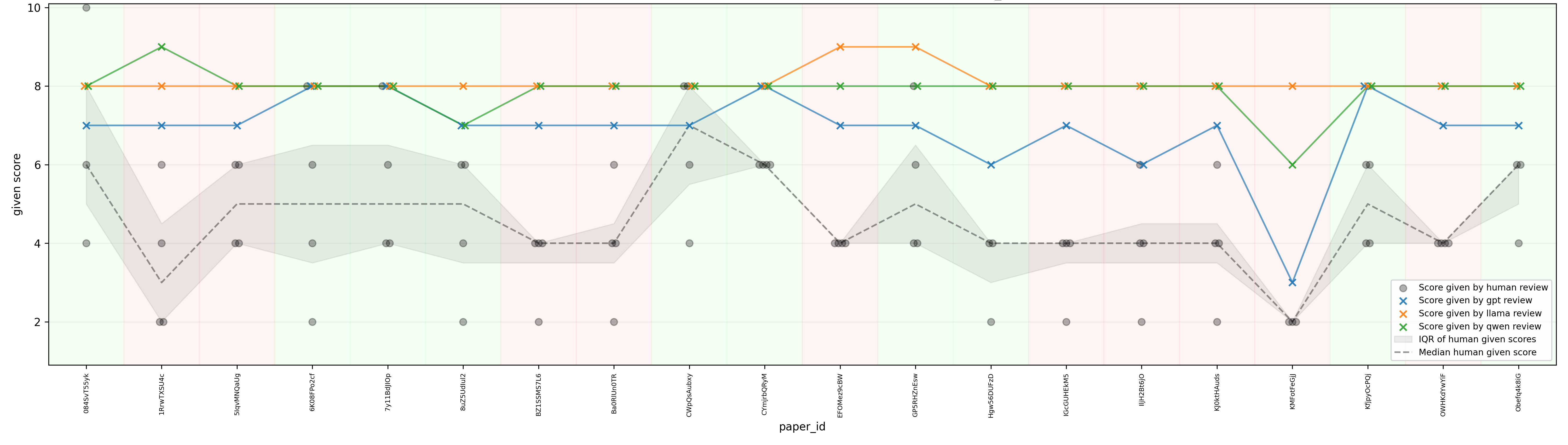}
    \caption{Scores assigned to individual ICLR submissions. Gray points show scores from individual human reviewers, while the colored lines show scores assigned by GPT-, Llama-, and Qwen-generated reviews. Green and red shading indicate accepted and rejected submissions, respectively.}
    \label{fig:givenScorePerPaper_ICLR} 
\end{figure*}

The main text reports score alignment using the mean human score for each
paper and notes that aggregation by the median yields the same conclusion.
\autoref{tab:given_score_correlation} provides the complete Pearson and
Spearman correlations for both aggregation methods.
\autoref{fig:givenScorePerPaper_ICLR} additionally visualizes the scores
assigned to each ICLR submission by individual human reviewers and the
three model-generated reviews, together with the final acceptance decision.

\begin{table}[ht]
\small
\centering
\begin{tabular}{@{}l
S S
S S@{}}
\toprule
& \multicolumn{2}{c}{\textbf{Human Mean}}
& \multicolumn{2}{c}{\textbf{Human Median}}\\
\cmidrule(lr){2-3}\cmidrule(lr){4-5}
\textbf{Model}
& \multicolumn{1}{c}{\makecell{\textbf{Pearson}\\$r$}}
& \multicolumn{1}{c}{\makecell{\textbf{Spearman}\\$\rho$}}
& \multicolumn{1}{c}{\makecell{\textbf{Pearson}\\$r$}}
& \multicolumn{1}{c}{\makecell{\textbf{Spearman}\\$\rho$}}\\
\midrule
GPT & {\bfseries $+0.62^{*}$} & {\bfseries $+0.60^{*}$} & {\bfseries $+0.55^{*}$} & {\bfseries $+0.51^{*}$}\\
Llama         & -0.01 & -0.02 & +0.05 & +0.08\\
Qwen          & +0.13 & +0.09 & +0.13 & +0.07\\
\bottomrule
\end{tabular}
\caption{Pearson and Spearman correlation between LLM-assigned scores and aggregated human review scores on ICLR ($n=50$ papers). Human scores are aggregated using either the mean or the median across reviewers. Asterisks denote statistically significant correlations ($^{*}p<0.001$, two-tailed).}
\label{tab:given_score_correlation}
\end{table}

\section{Per-Paper GranuScore Results for Nature Communications}
\label{app:granuscore-results}

\begin{figure*}[ht]
    \centering
    \includegraphics[width=\textwidth]{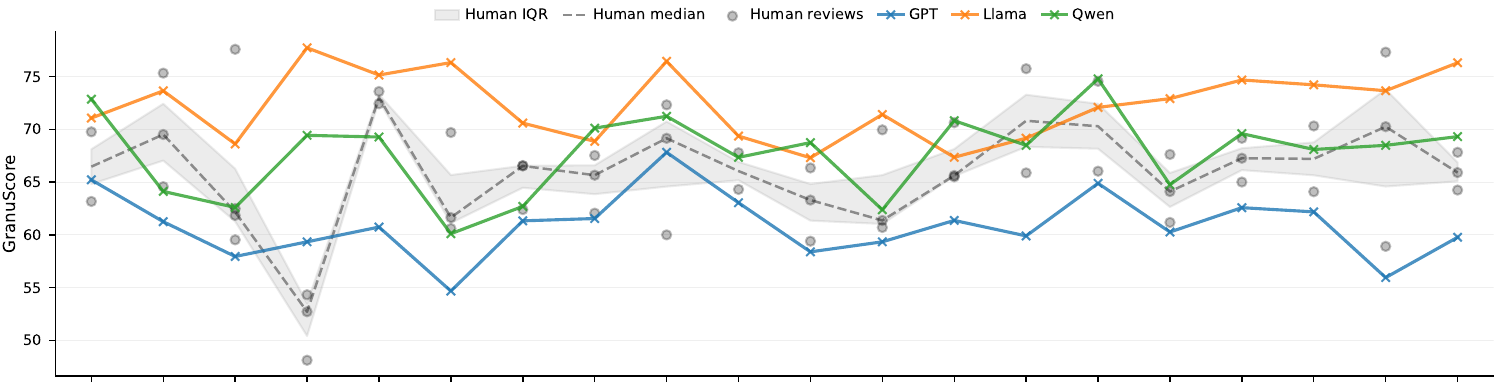}
    \caption{Per-paper GranuScore for Nature Communications reviews. Gray points show human reviews, with their median and interquartile range indicated by the dashed line and shaded band; colored lines show generated reviews.}
    \label{fig:granuscore_per_paper_NatureCom}
\end{figure*}

The main text reports the aggregate GranuScore results for both venues and
shows the per-paper distributions for ICLR. \autoref{fig:granuscore_per_paper_NatureCom}
provides the corresponding visualization for \textit{Nature Communications}.

\section{LLM-as-a-Judge Results}
\label{app:llm-as-a-judge-results}
\begin{figure*}[ht]
    \centering
    \includegraphics[width=1\textwidth]{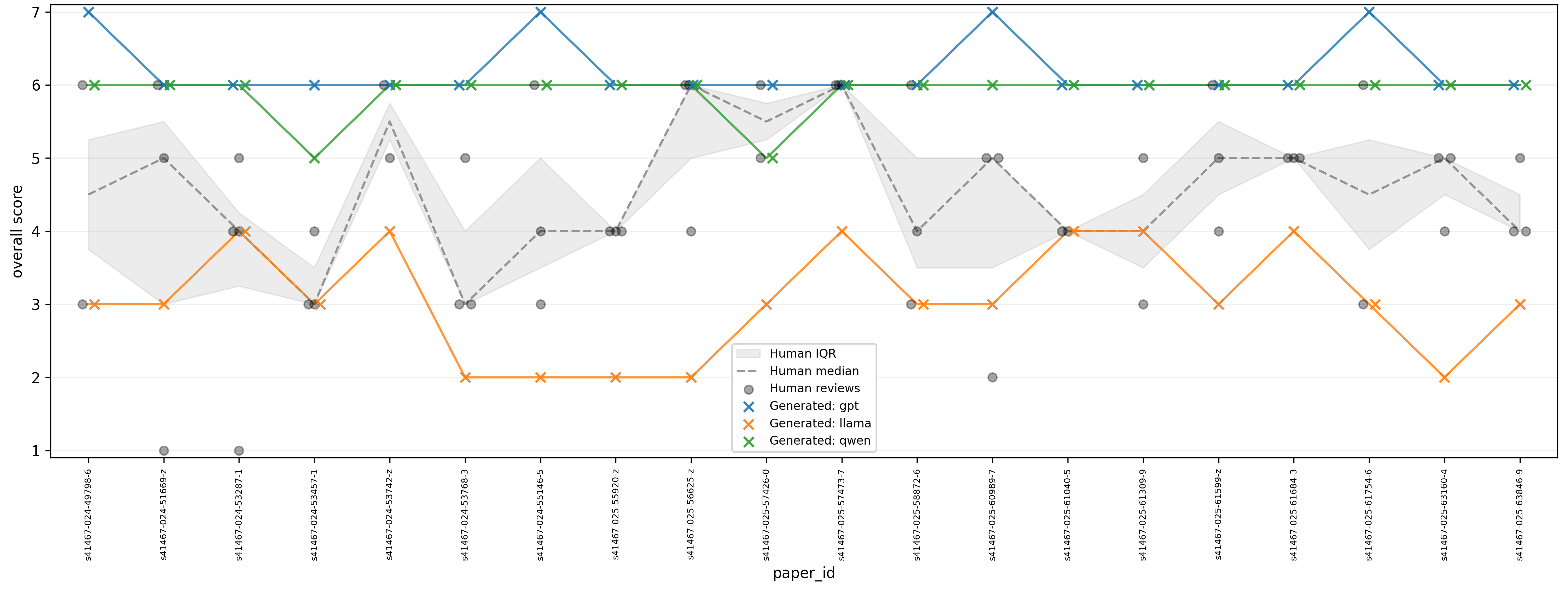}
    \caption{Overall score given by LLM-as-a-Judge for the peer-reviews of Nature Communication papers.} \label{fig:overall_score_NatureCom} 
\end{figure*}

\begin{figure}[ht]
    \centering
    \includegraphics[width=1\textwidth]{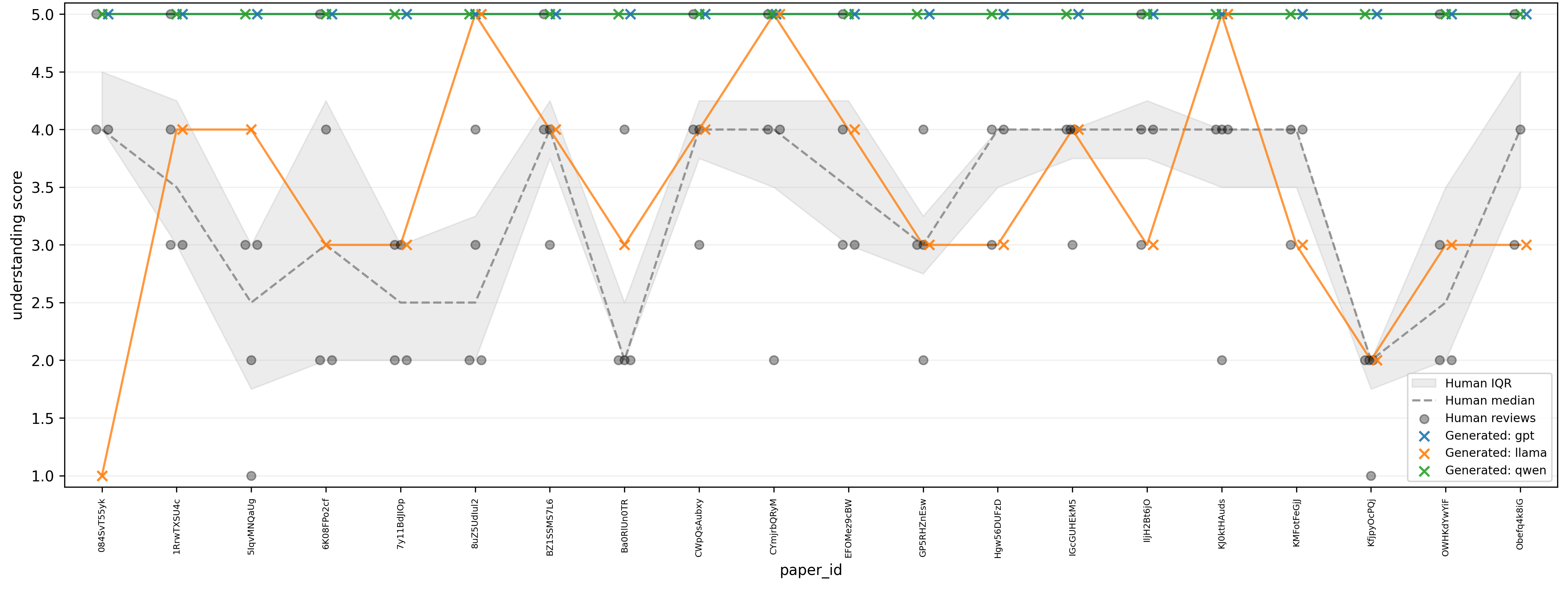}
    \caption{Understanding score given by LLM-as-a-Judge for the peer-reviews of ICLR papers.} \label{fig:understanding_score_ICLR} 
\end{figure}

\begin{figure}[ht]
    \centering
    \includegraphics[width=1\textwidth]{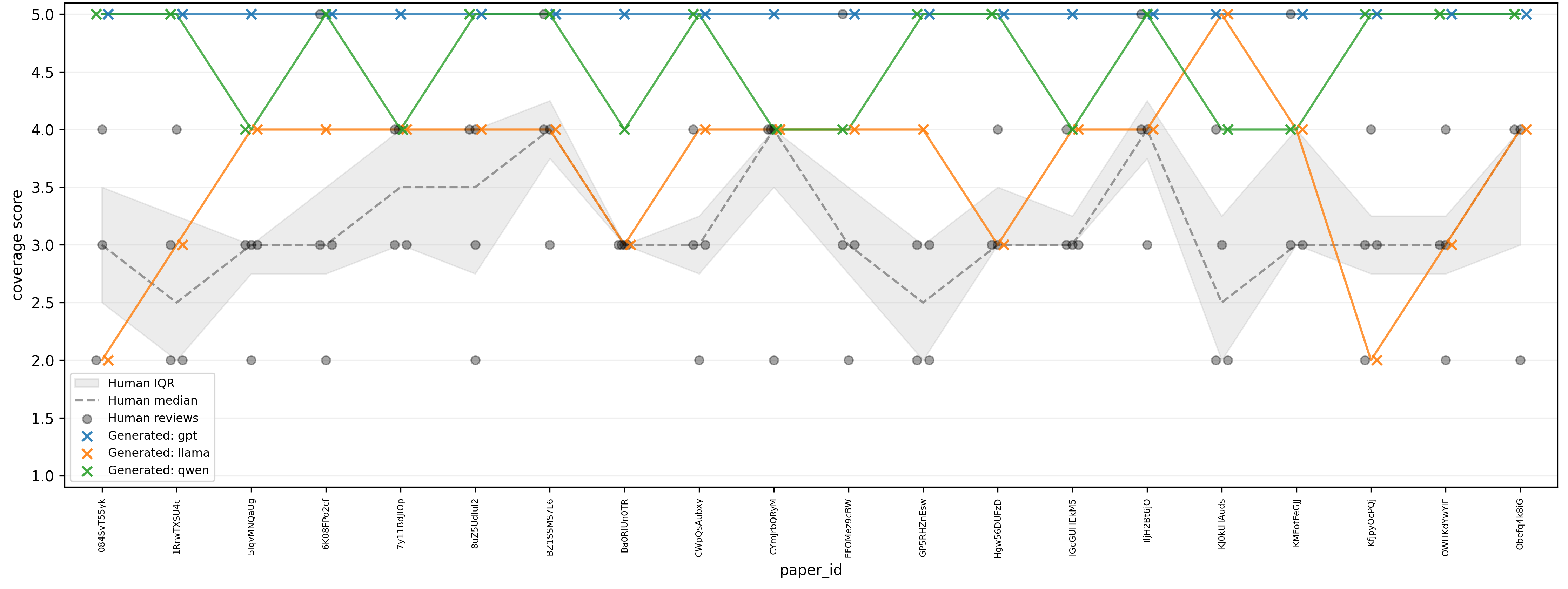}
    \caption{Coverage score given by LLM-as-a-Judge for the peer-reviews of ICLR papers.} \label{fig:coverage_score_ICLR} 
\end{figure}

\begin{figure}[ht]
    \centering
    \includegraphics[width=1\textwidth]{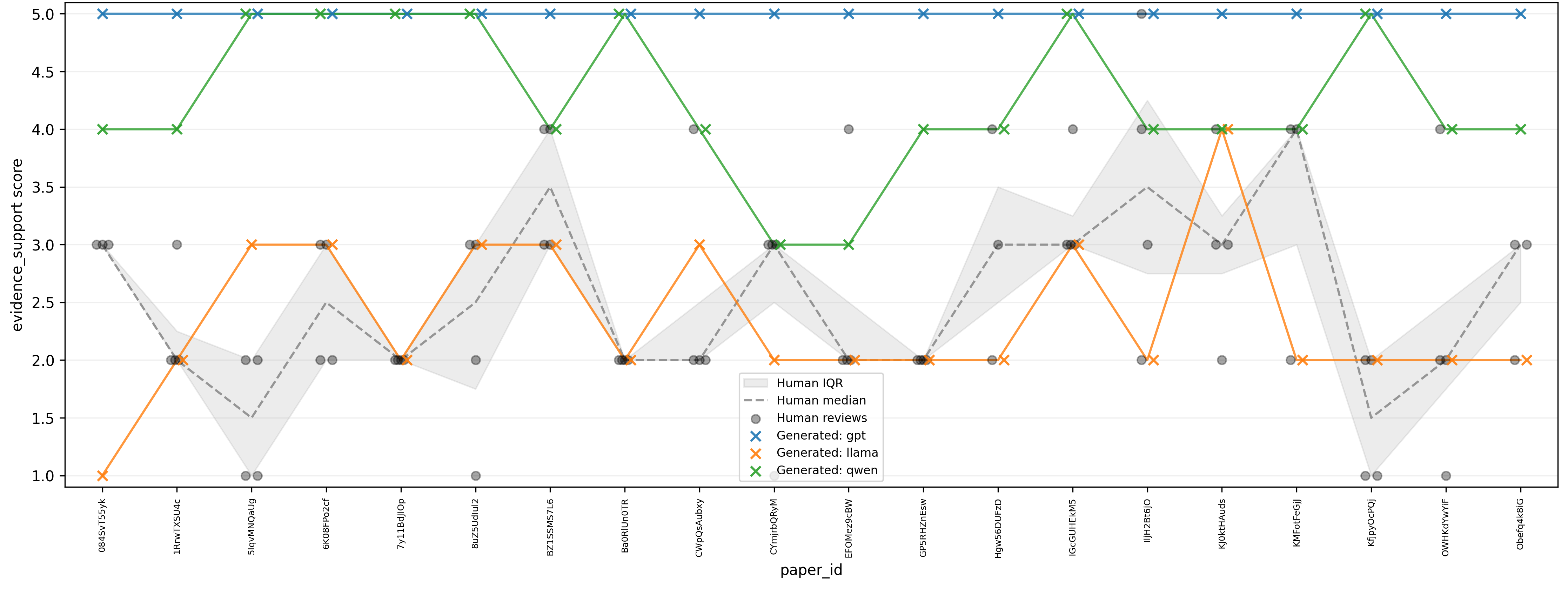}
    \caption{Evidence support score given by LLM-as-a-Judge for the peer-reviews of ICLR papers.} \label{fig:evidence_support_score_ICLR} 
\end{figure}

\begin{figure}[ht]
    \centering
    \includegraphics[width=1\textwidth]{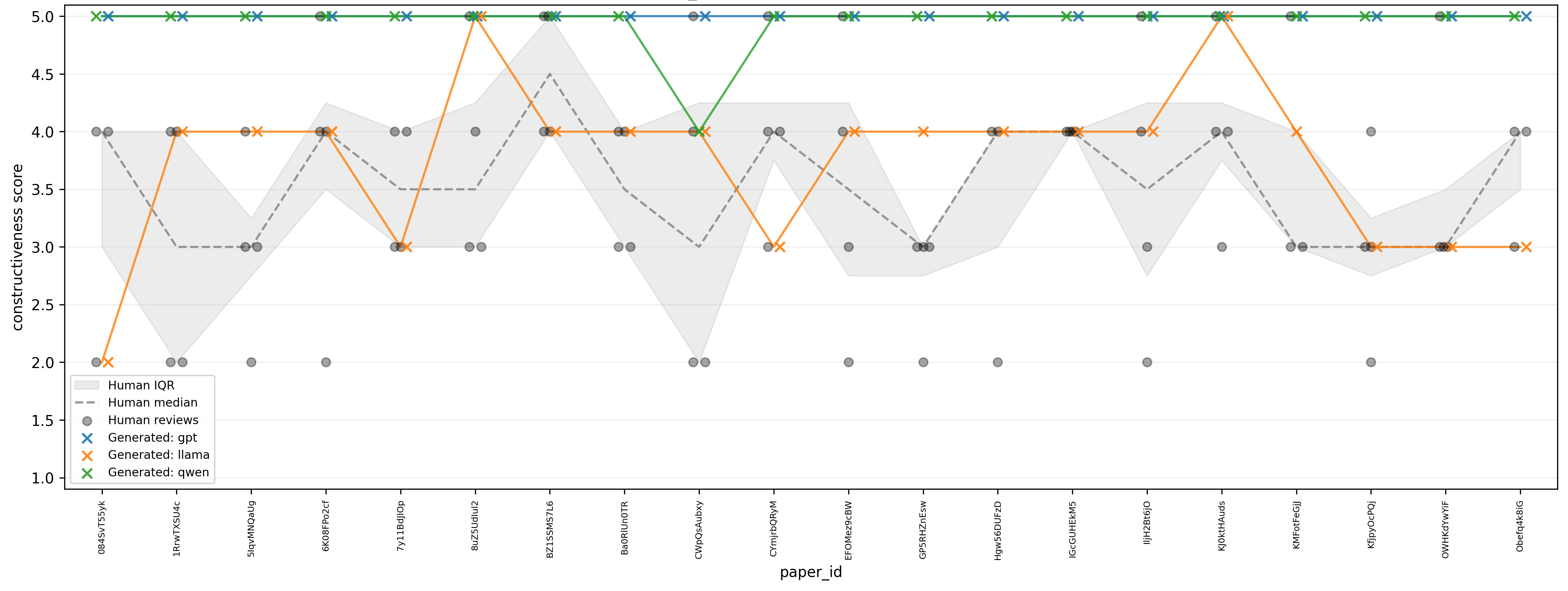}
    \caption{Constructiveness score given by LLM-as-a-Judge for the peer-reviews of ICLR papers.} \label{fig:constructiveness_score_ICLR} 
\end{figure}

\begin{figure}[ht]
    \centering
    \includegraphics[width=1\textwidth]{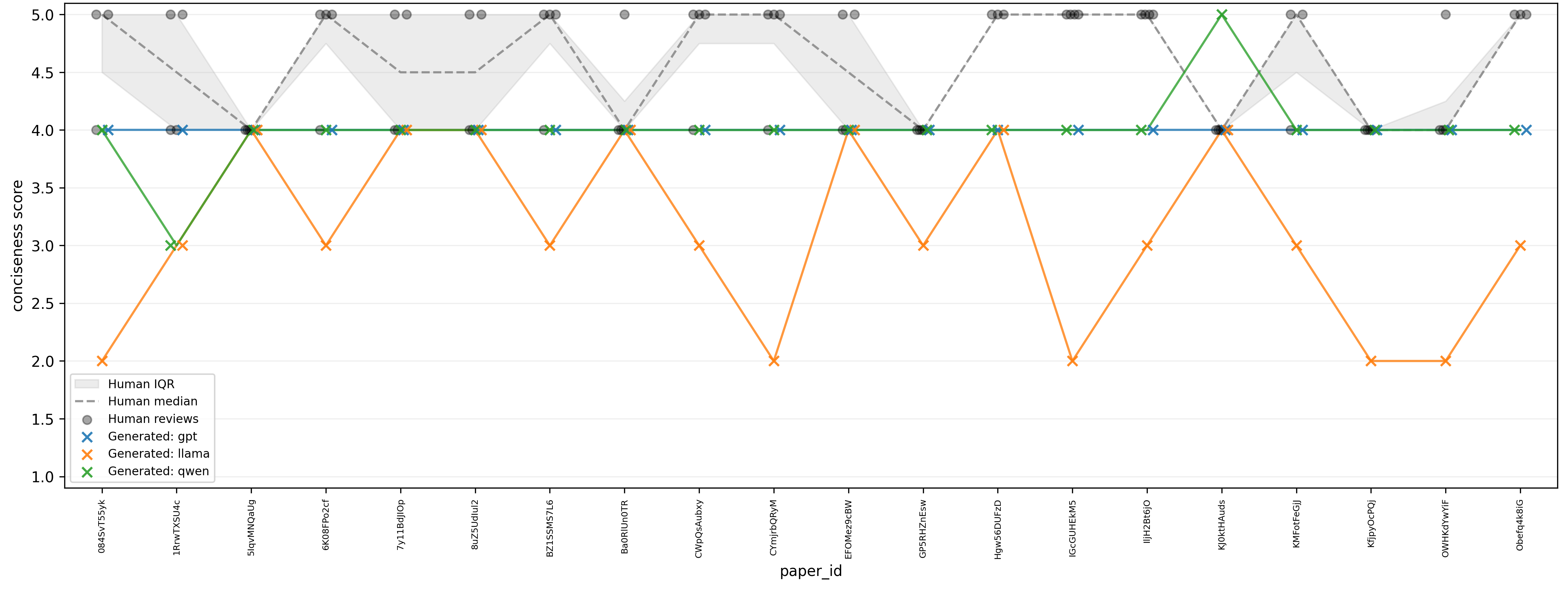}
    \caption{Conciseness score given by LLM-as-a-Judge for the peer-reviews of ICLR papers.} \label{fig:conciseness_score_ICLR} 
\end{figure}

\begin{figure}[ht]
    \centering
    \includegraphics[width=1\textwidth]{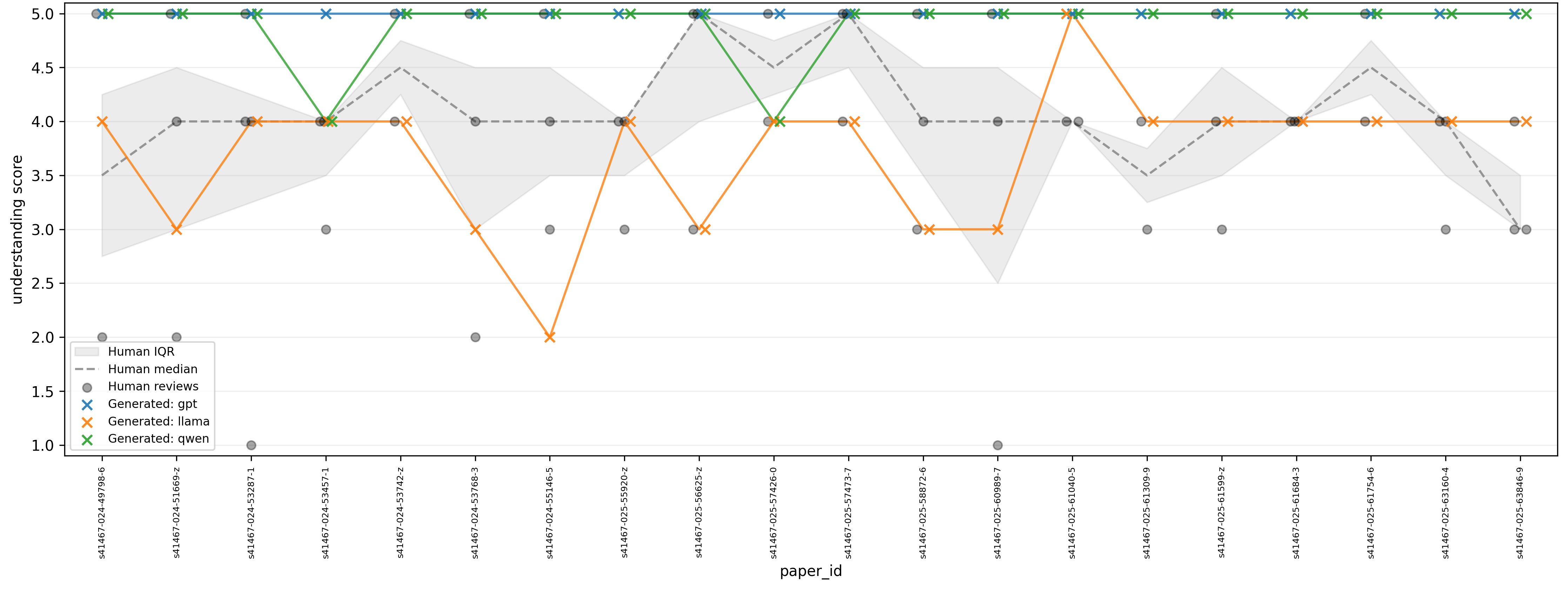}
    \caption{Understanding score given by LLM-as-a-Judge for the peer-reviews of Nature Communications papers.} \label{fig:understanding_score_NatureCom} 
\end{figure}

\begin{figure}[ht]
    \centering
    \includegraphics[width=1\textwidth]{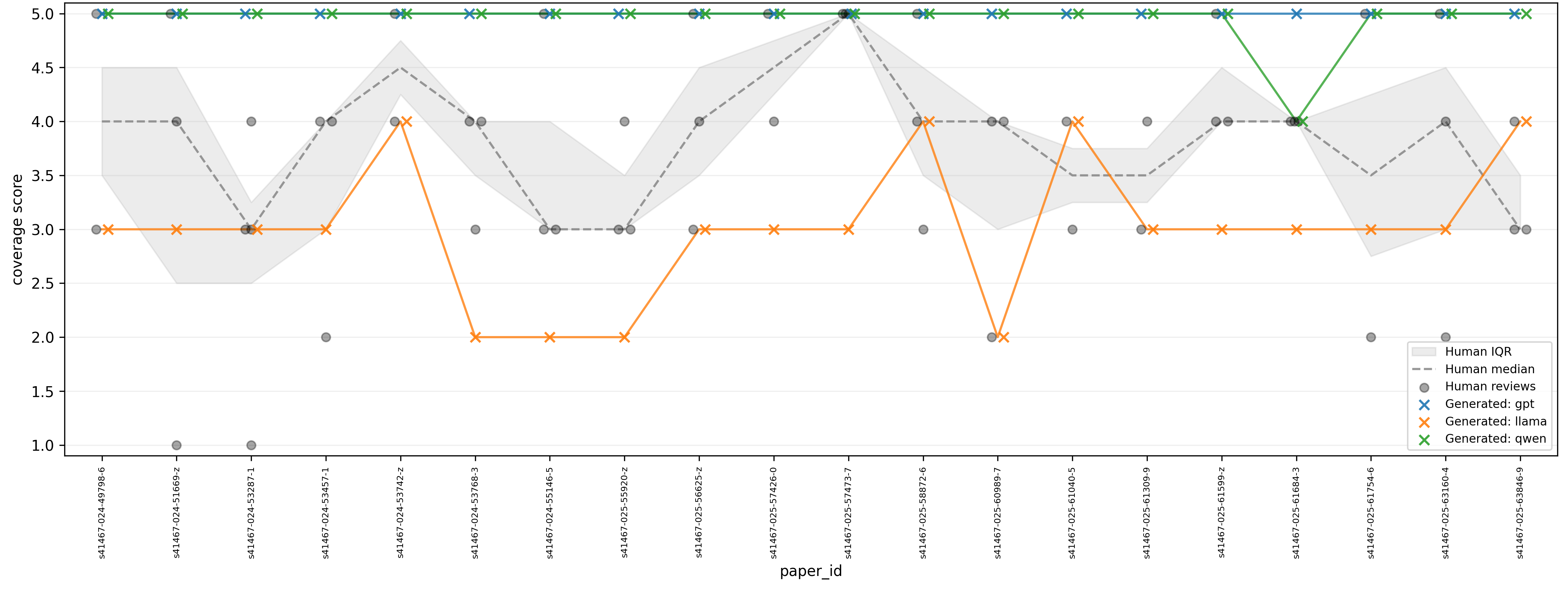}
    \caption{Coverage score given by LLM-as-a-Judge for the peer-reviews of Nature Communications papers.} \label{fig:coverage_score_NatureCom} 
\end{figure}

\begin{figure}[ht]
    \centering
    \includegraphics[width=1\textwidth]{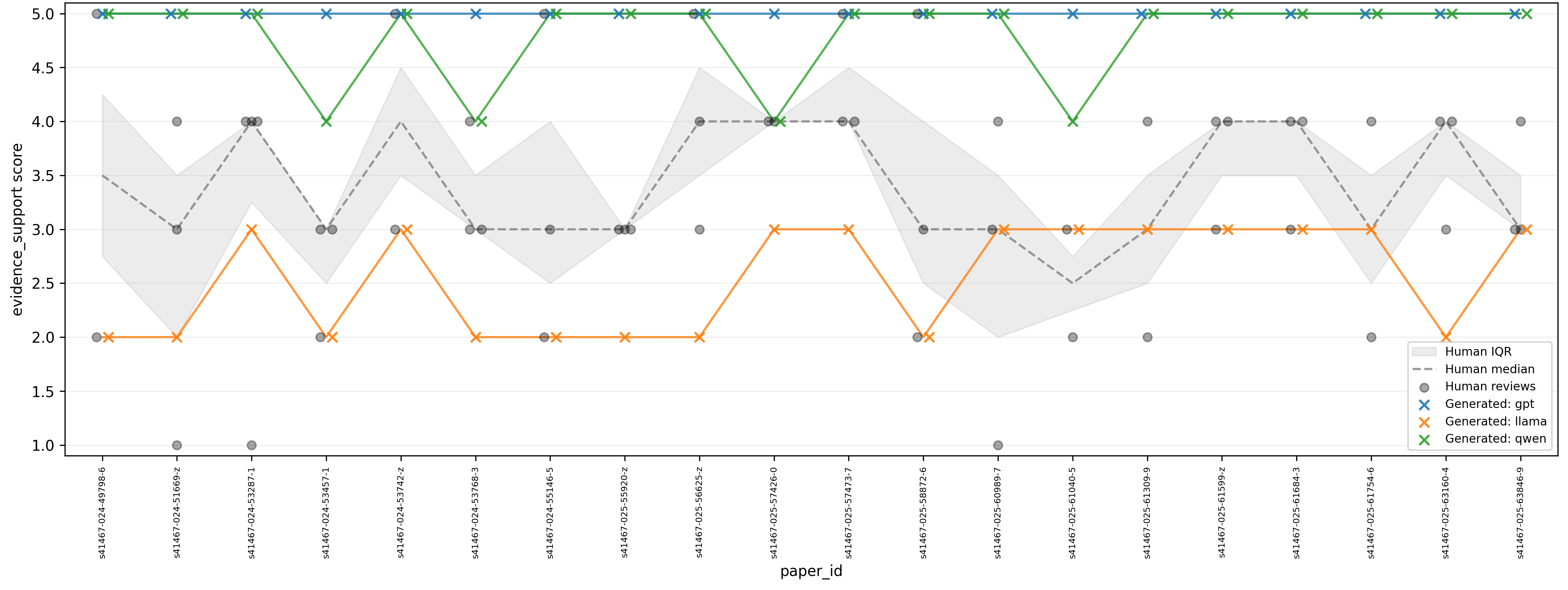}
    \caption{Evidence support score given by LLM-as-a-Judge for the peer-reviews of Nature Communications papers.} \label{fig:evidence_support_score_NatureCom} 
\end{figure}

\begin{figure}[ht]
    \centering
    \includegraphics[width=1\textwidth]{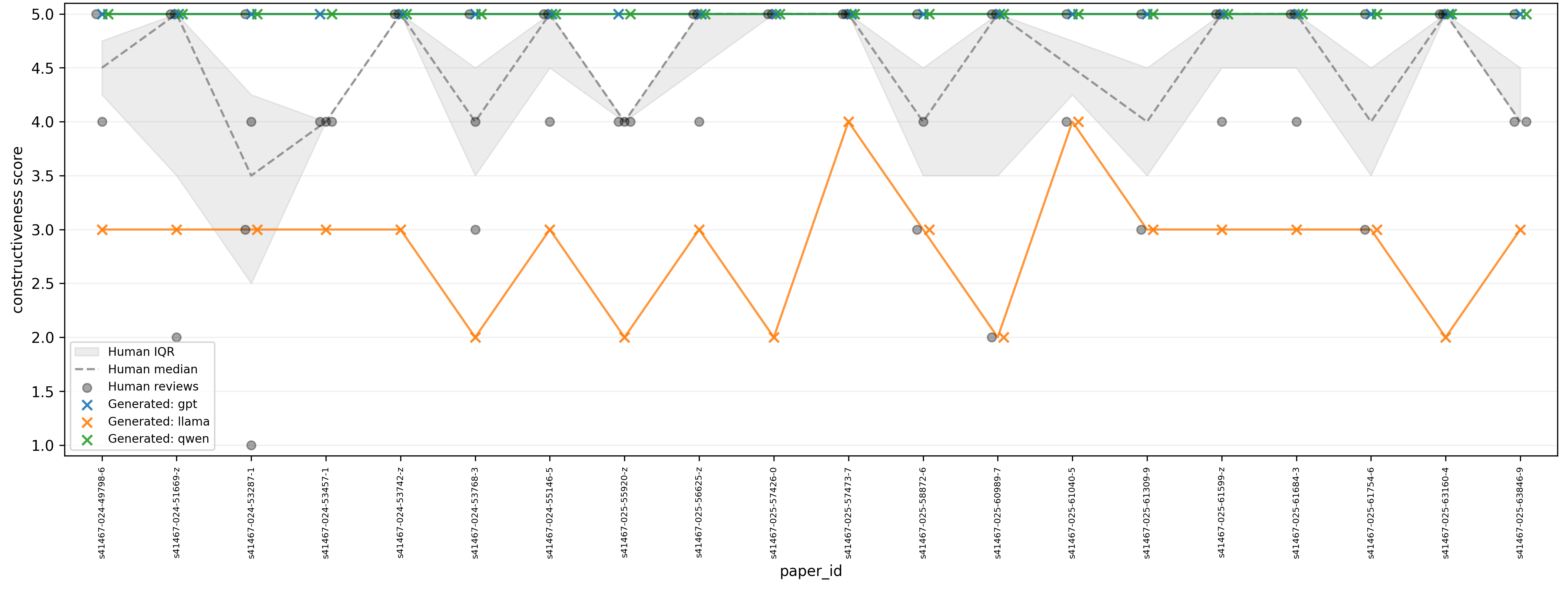}
    \caption{Constructiveness score given by LLM-as-a-Judge for the peer-reviews of Nature Communications papers.} \label{fig:constructiveness_score_NatureCom} 
\end{figure}

\begin{figure}[ht]
    \centering
    \includegraphics[width=1\textwidth]{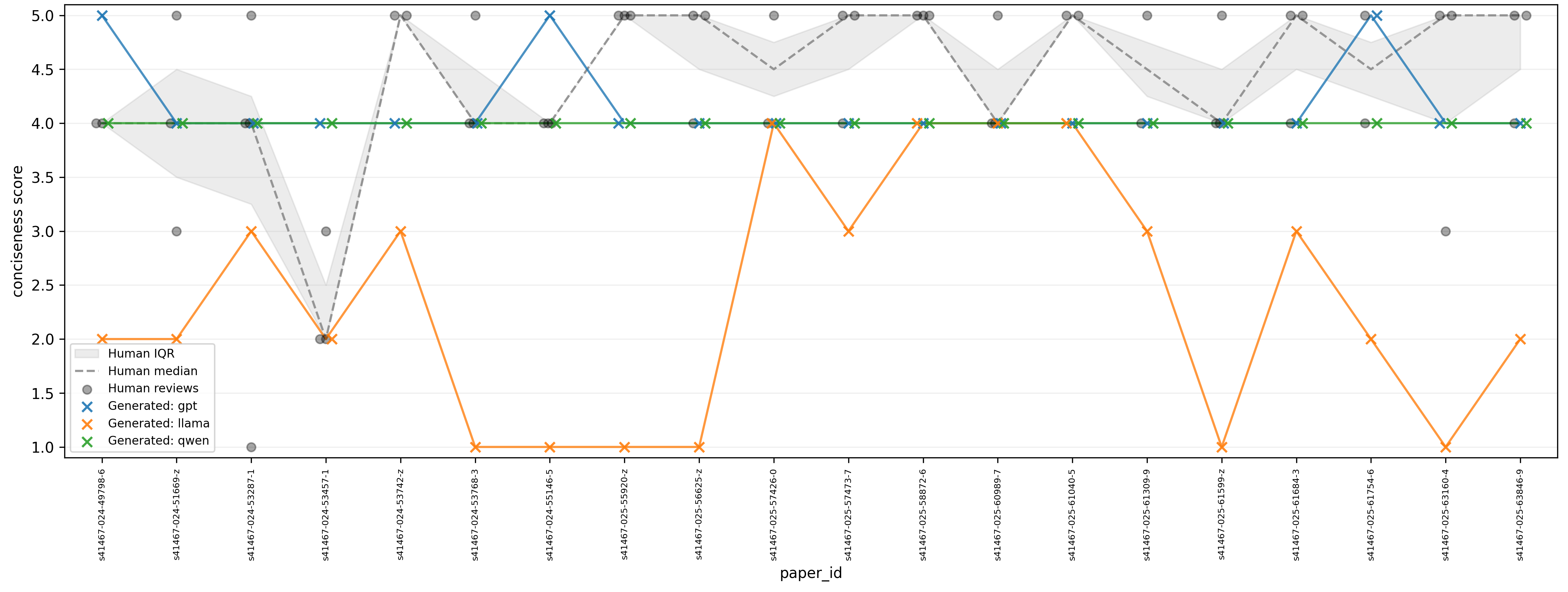}
    \caption{Conciseness score given by LLM-as-a-Judge for the peer-reviews of Nature Communications papers.} \label{fig:conciseness_score_NatureCom} 
\end{figure}

This section supplements the aggregate LLM-as-a-judge results reported in
the main text with detailed score distributions and results stratified by
ICLR submission decision. \autoref{fig:overall_score_NatureCom} shows the
overall-score distribution for \textit{Nature Communications}, complementing
the ICLR distribution shown in the main text. Figures
\ref{fig:understanding_score_ICLR}--\ref{fig:conciseness_score_ICLR} and
\ref{fig:understanding_score_NatureCom}--\ref{fig:conciseness_score_NatureCom}
show the corresponding per-paper distributions for understanding, coverage,
evidence support, constructiveness, and conciseness for ICLR and
\textit{Nature Communications}, respectively.

\autoref{tab:llmjudge_by_status_full} further compares judged review quality
for accepted and rejected ICLR submissions. Across all review sources and
dimensions, the group means differ by at most $0.37$ points, with most
differences being smaller. Judged review quality is therefore largely
independent of the paper's eventual acceptance decision.

\begin{table*}[ht]
\small
\centering
\begin{tabular}{ll
S[table-format=1.1(1.1)] S[table-format=1.1(1.1)] S[table-format=1.1(1.1)]
S[table-format=1.1(1.1)] S[table-format=1.1(1.1)] S[table-format=1.1(1.1)]}
\toprule
\textbf{Source} & \textbf{Decision} &
{\textbf{Overall}} & {\textbf{Underst.}} & {\textbf{Coverage}} &
{\textbf{Evid.\ Sup.}} & {\textbf{Constr.}} & {\textbf{Concis.}} \\
\midrule
\multirow{2}{*}{Human} & Accepted & 3.5(1.0) & 3.2(1.0) & 3.1(0.9) & 2.3(0.7) & 3.5(0.8) & 4.5(0.5) \\
                       & Rejected & 3.6(1.1) & 3.2(1.0) & 3.3(0.9) & 2.6(0.9) & 3.6(1.0) & 4.4(0.6) \\
\midrule
\multirow{2}{*}{GPT}   & Accepted & 6.9(0.4) & 5.0(0.0) & 5.0(0.0) & 5.0(0.0) & 5.0(0.0) & 4.0(0.0) \\
                       & Rejected & 6.7(0.5) & 5.0(0.0) & 5.0(0.0) & 5.0(0.0) & 5.0(0.0) & 4.0(0.0) \\
\midrule
\multirow{2}{*}{Llama} & Accepted & 3.3(0.9) & 3.2(1.0) & 3.5(0.8) & 2.2(0.6) & 3.5(0.8) & 2.8(0.8) \\
                       & Rejected & 3.7(0.7) & 3.5(0.7) & 3.8(0.5) & 2.5(0.7) & 3.9(0.6) & 3.1(0.7) \\
\midrule
\multirow{2}{*}{Qwen}  & Accepted & 5.8(0.4) & 5.0(0.0) & 4.7(0.5) & 4.5(0.6) & 5.0(0.2) & 3.9(0.4) \\
                       & Rejected & 5.8(0.5) & 4.9(0.4) & 4.6(0.5) & 4.3(0.8) & 5.0(0.0) & 4.0(0.3) \\
\bottomrule
\end{tabular}
\caption{LLM-as-a-judge scores for reviews of accepted and rejected ICLR
submissions (mean $\pm$ std.\ across papers). Overall quality is rated on
a 1--7 scale; all other dimensions are rated on a 1--5 scale. Across all
review sources and dimensions, the group means differ by at most $0.37$
points.}
\label{tab:llmjudge_by_status_full}
\end{table*}

\section{Additional Qualitative Analysis}
\label{app:qualitative-analysis}

\subsection{Positive Bias in Generated Reviews}
\label{app:pos-bias}
Across the 31 papers, $30/31$ Llama reviews contain at least one of the phrases \textit{``highly recommend''}, \textit{``suitable for publication''}, \textit{``make a significant impact''}, \textit{``valuable contribution''}, \textit{``significant contribution''}, or \textit{``well-written''}; the phrase \textit{``well-written''} alone appears in $27/31$ reviews and
\textit{``significant contribution''} in $20/31$. The same phrases appear in only $3/31$ GPT reviews and $22/31$ Qwen reviews, and in $20/96$ of the
individual human reviewer remarks we extracted for Nature Communications. On ICLR, $47/50$ Llama reviews contain at least one such phrase, compared with $16/50$ for GPT, $21/50$ for Qwen, and $44/192$ of the corresponding human reviews.

\subsection{Evidence Grounding}
\label{app:evidence-grounding}
A peer review is more useful when it points authors to specific locations in
the manuscript. We counted references to ``Figure'', ``Table'', ``Section'',
and ``Equation'' followed by a digit.

On ICLR, GPT produces an average of $29.8$ figure, $34.5$ table, $42.0$
section, and $6.7$ equation references per review. Qwen produces
$10.8 / 8.8 / 8.3 / 0.5$, whereas Llama produces almost none
($0.3 / 0.1 / 0.0 / 0.0$). Human reviews in the same dataset average only
$0.5 / 0.5 / 0.2 / 0.1$ references.

Thus, GPT substantially exceeds the human baseline in explicit manuscript
references, whereas Llama rarely provides such grounding.

\subsection{Generation Failures (LLama)}
In $5/31$ cases, the Llama review for Nature Communications papers degenerates into a run of short, single-sentence praises without any concrete evidence. The most extreme case is \texttt{s41467-\allowbreak 025-66620-\allowbreak z}, where the closing of the review consists of $64$ consecutive short praise sentences:
\begin{quote}%\small
\textit{``The manuscript is suitable for publication. The study will make a
significant contribution to the field. The study has been well-designed and
well-executed. The study has significant clinical implications. The study will improve clinical practice. \ldots The study is a valuable addition to the field.''}
\end{quote}
The same pattern appears in \texttt{s41467-\allowbreak 024-\allowbreak 53768-\allowbreak 3}, \texttt{s41467-\allowbreak 025-\allowbreak 61754-\allowbreak 6}, \texttt{s41467-\allowbreak 025-\allowbreak 64227-\allowbreak y}, and \texttt{s41467-\allowbreak 025-\allowbreak 61599-\allowbreak z}.
It likely accounts for Llama's low conciseness score ($2.3 \pm 1.1$ on Nature Communications, vs.\ $4.4 \pm 0.8$ for human
reviewers) in Figure \ref{fig:conciseness_score_NatureCom}.

\subsection{Generic Critiques (LLama)}
We furthermore scanned the ICLR Weaknesses (Q6) and Suggestions (Q7) fields
for stock phrasings that could be applied to almost any ML paper, such as \textit{``more detailed/comprehensive analysis''}, \textit{``hyperparameter sensitivity''}, \textit{``comparison with more recent work''}, \textit{``real-world datasets/scenarios''}, and \textit{``computational cost''}. Llama produces one of these stock
phrasings in between $9$ and $37$ of $50$ papers per phrase, Qwen in $2$ to $24$, and GPT in $0$ to $8$. The corresponding counts over the $192$ human reviews never exceed $14$. A representative Llama weakness reads
\textit{``Limited Analysis of Hyperparameter Sensitivity: The paper could benefit from a more comprehensive analysis of hyperparameter sensitivity\ldots A more detailed analysis would help readers understand the robustness of the method''}, a complaint that could be copy-pasted onto essentially any ML submission.

\subsection{Ethics Field Over-Production (Qwen)}
The ICLR ethics field (Q8) explicitly enumerates \textit{``No ethics review needed''} as a valid answer. GPT and Llama use this short answer in the majority of papers (median answer length $450$ and $23$ characters, respectively). Qwen, in contrast, produces a median of $1{,}089$ characters and a maximum of $1{,}747$ characters per ethics field, restating ICLR's ethics policy and reasoning at length about why no ethics review is needed. This inflates Qwen's review length without adding informative content for either authors or area chairs.

\section{Survey of AI Review Policies}
\label{app:survey-tables}
\autoref{tab:survey_astar} and \autoref{tab:appendix_medical} provide the
complete venue-level classifications for the surveyed A$^*$ AI/NLP venues
and medical journals, respectively.

\begin{table}[ht]
\centering
\small
\begin{tabular}{llll}
\toprule
\textbf{Venue} & \textbf{Rank} & \textbf{Policy} & \textbf{Sub-tags} \\
\midrule
AAAI     & A$^*$ & Active  & \\
AAMAS    & A$^*$ & Prohib. & \\
ACL      & A$^*$ & Partial & \texttt{polish\_ok} \\
ACMMM    & A$^*$ & Partial & \texttt{readability\_ok; no\_manuscript\_upload} \\
COLT     & A$^*$ & Partial & \texttt{no\_manuscript\_upload} \\
CVPR     & A$^*$ & Prohib. & \\
EC       & A$^*$ & Partial & \texttt{polish\_ok; readability\_ok; no\_manuscript\_upload} \\
ECCV     & A$^*$ & Partial & \texttt{polish\_ok} \\
EMNLP    & A$^*$ & Partial & \texttt{polish\_ok} \\
ICAPS    & A$^*$ & Partial & \texttt{polish\_ok} \\
ICCV     & A$^*$ & Prohib. & \\
ICDE     & A$^*$ & Prohib. & \\
ICDM     & A$^*$ & Prohib. & \\
ICLR     & A$^*$ & Partial & \texttt{disclosure\_required} \\
ICML     & A$^*$ & Partial & \\
ICRA     & A$^*$ & Prohib. & \\
IJCAI    & A$^*$ & Partial & \texttt{polish\_ok} \\
KDD      & A$^*$ & Partial & \texttt{no\_manuscript\_upload} \\
KR       & A$^*$ & None    & \\
NeurIPS  & A$^*$ & Partial & \texttt{polish\_ok; no\_manuscript\_upload} \\
PODS     & A$^*$ & Partial & \texttt{readability\_ok} \\
SIGGRAPH & A$^*$ & Prohib. & \\
SIGIR    & A$^*$ & Partial & \texttt{readability\_ok} \\
SIGMOD   & A$^*$ & Partial & \texttt{readability\_ok} \\
VLDB     & A$^*$ & None    & \\
WWW      & A$^*$ & Partial & \texttt{readability\_ok} \\
\bottomrule
\end{tabular}
\caption{Reviewer-AI policy for the A$^*$ AI/NLP venues ($n=26$). Policy: Prohib.\ = prohibited; Partial = partially allowed; Active = actively used; None = no policy. Sub-tags apply only to Partial venues; an empty cell for a Partial venue (ICML) means the policy states only what is banned.}
\label{tab:survey_astar}
\end{table}

\begin{table}[ht]
\centering
\small
\begin{tabular}{lrlll}
\toprule
\textbf{Journal} & \textbf{JIF} & \textbf{Subcat.} & \textbf{Policy} & \textbf{Sub-tags} \\
\midrule
CA-CANCER J CLIN     & 232.4 & Oncology  & Partial & \texttt{no\_manuscript\_upload} \\
LANCET               &  88.5 & Gen.\ Int. & Prohib. & \\
NAT REV CLIN ONCOL   &  83.2 & Oncology  & Partial & \texttt{no\_manuscript\_upload} \\
NEW ENGL J MED       &  78.5 & Gen.\ Int. & Partial & \texttt{disclosure\_required} \\
NAT REV CANCER       &  66.8 & Oncology  & Partial & \texttt{no\_manuscript\_upload} \\
ANN ONCOL            &  65.4 & Oncology  & Prohib. & \\
NAT REV DIS PRIMERS  &  60.6 & Gen.\ Int. & Partial & \texttt{no\_manuscript\_upload} \\
JAMA-J AM MED ASSOC  &  55.0 & Gen.\ Int. & Prohib. & \\
LANCET NEUROL        &  45.5 & Neuro.    & Prohib. & \\
CANCER CELL          &  44.5 & Oncology  & Prohib. & \\
NAT REV CARDIOL      &  44.2 & Cardio.   & Partial & \texttt{no\_manuscript\_upload} \\
J CLIN ONCOL         &  43.4 & Oncology  & Prohib. & \\
BMJ-BRIT MED J       &  43.0 & Gen.\ Int. & Partial & \texttt{polish\_ok; disclosure\_required} \\
J HEMATOL ONCOL      &  40.4 & Oncology  & Partial & \texttt{no\_manuscript\_upload} \\
CIRCULATION          &  38.7 & Cardio.   & Prohib. & \\
LANCET ONCOL         &  35.9 & Oncology  & Prohib. & \\
EUR HEART J          &  35.7 & Cardio.   & Partial & \texttt{no\_manuscript\_upload} \\
NAT REV NEUROL       &  33.1 & Neuro.    & Partial & \texttt{no\_manuscript\_upload} \\
LANCET DIGIT HEALTH  &  24.1 & Gen.\ Int. & Prohib. & \\
JAMA INTERN MED      &  23.3 & Gen.\ Int. & Prohib. & \\
MILITARY MED RES     &  22.9 & Gen.\ Int. & Partial & \texttt{no\_manuscript\_upload} \\
J AM COLL CARDIOL    &  22.3 & Cardio.   & Prohib. & \\
JAMA NEUROL          &  21.4 & Neuro.    & Prohib. & \\
CIRC RES             &  16.2 & Cardio.   & Prohib. & \\
JACC-CARDIOVASC IMAG &  15.2 & Cardio.   & Prohib. & \\
RADIOLOGY            &  15.2 & Radiol.   & Partial & \texttt{no\_manuscript\_upload} \\
NPJ DIGIT MED        &  15.1 & Health    & Partial & \texttt{no\_manuscript\_upload} \\
JAMA CARDIOL         &  14.1 & Cardio.   & Prohib. & \\
IMPLEMENT SCI        &  13.4 & Health    & Partial & \texttt{no\_manuscript\_upload} \\
JACC-CARDIOONCOL     &  13.4 & Cardio.   & Prohib. & \\
NEURO-ONCOLOGY       &  13.4 & Neuro.    & Partial & \texttt{disclosure\_required} \\
RADIOL-ARTIF INTELL  &  13.2 & Radiol.   & Partial & \texttt{no\_manuscript\_upload} \\
LANCET REG HEALTH-EU &  13.0 & Health    & Prohib. & \\
MED IMAGE ANAL       &  11.8 & Radiol.   & Prohib. & \\
BRAIN                &  11.7 & Neuro.    & None    & \\
JAMA-HEALTH FORUM    &  11.3 & Health    & Prohib. & \\
ALZHEIMERS DEMENT    &  11.1 & Neuro.    & Partial & \texttt{no\_manuscript\_upload} \\
IEEE T MED IMAGING   &   9.8 & Radiol.   & Active  & \\
SLEEP MED REV        &   9.7 & Neuro.    & Prohib. & \\
CLIN NUCL MED        &   9.6 & Radiol.   & Prohib. & \\
ACTA NEUROPATHOL     &   9.3 & Neuro.    & Partial & \texttt{no\_manuscript\_upload} \\
J NUCL MED           &   9.1 & Radiol.   & Prohib. & \\
DIAGN INTERV IMAG    &   8.1 & Radiol.   & Prohib. & \\
HEALTH AFFAIR        &   8.1 & Health    & Prohib. & \\
LANCET REG HEALTH-W  &   8.1 & Health    & Prohib. & \\
INVEST RADIOL        &   8.0 & Radiol.   & Prohib. & \\
PLOS DIGIT HEALTH    &   7.7 & Health    & Prohib. & \\
LANCET REG HEALTH-AM &   7.6 & Health    & Prohib. & \\
\bottomrule
\end{tabular}
\caption{All medical journals in the survey ($n=48$), sorted by 2024 Journal Impact Factor. Subcategory abbreviations: Gen.\ Int.\ = Medicine, General \& Internal; Cardio.\ = Cardiac \& Cardiovascular Systems; Neuro.\ = Clinical Neurology; Radiol.\ = Radiology, Nuclear Medicine \& Medical Imaging; Health = Health Care Sciences \& Services.}
\label{tab:appendix_medical}
\end{table}

\end{document}